\documentclass[reqno, eucal]{amsart}

\usepackage[mathscr]{eucal} 
\usepackage[dvips]{graphicx}
\usepackage[dvips]{color}
\usepackage{amsmath,amsfonts,amssymb,amsthm,amscd}
\usepackage{epic}
\usepackage{eepic}
\usepackage{longtable}
\usepackage{array}

\theoremstyle{definition}

\newcommand{\Z}{{\mathbb Z}}

\newcommand{\C}{{\mathbb C}}

\newcommand{\I}{{\mathrm i}}

\newcommand{\ok}{{\rm{\bf k}}}
\newcommand{\OK}{{\rm{\bf K}}}
\newcommand{\am}{{\rm{\bf a}}^{\!-} }
\newcommand{\ap}{{\rm{\bf a}}^{\!+} }
\newcommand{\apm}{{\rm{\bf a}}^{\!\pm} }
\newcommand{\amp}{{\rm{\bf a}}^{\!\mp} }
\newcommand{\Am}{{\rm{\bf A}}^{\!-} }
\newcommand{\Ap}{{\rm{\bf A}}^{\!+} }
\newcommand{\Apm}{{\rm{\bf A}}^{\!\pm} }
\newcommand{\Amp}{{\rm{\bf A}}^{\!\mp} }
\newcommand{\ichi}{1}
\newcommand{\hf}{{\scriptstyle \frac{1}{2}}}
\newcommand{\si}{{\scriptstyle 1}}
\newcommand{\sz}{{\scriptstyle 0}}
\newcommand{\Rm}{\mathscr{R}}
\newcommand{\Km}{\mathscr{K}}
\newcommand{\alb}{\boldsymbol{\alpha}}
\newcommand{\beb}{\boldsymbol{\beta}}
\newcommand{\gab}{\boldsymbol{\gamma}}
\newcommand{\deb}{\boldsymbol{\delta}}

\begin{document}

\title[Matrix product solutions 
to the reflection equation]{Matrix product solutions 
to the reflection equation \\
from three dimensional integrability}

\author{Atsuo Kuniba}
\email{atsuo.s.kuniba@gmail.com}
\address{Institute of Physics, 
University of Tokyo, Komaba, Tokyo 153-8902, Japan}

\author{Vincent Pasquier}
\email{vincent.pasquier@ipht.fr}
\address{Institut de Physique Th\'eorique, 
Universit\'e Paris Saclay, CEA, CNRS, F-91191 
Gif-sur-Yvette, France}

\maketitle

\vspace{0.5cm}
\begin{center}{\bf Abstract}
\end{center}

We formulate a quantized reflection equation in which 
$q$-boson valued $L$ and $K$ matrices
satisfy the reflection equation up to conjugation by 
a solution to the Isaev-Kulish 3D reflection equation. 
By forming its $n$-concatenation along the $q$-boson Fock space
followed by suitable reductions, 
we construct families of solutions to 
the reflection equation 
in a matrix product form connected to the 3D integrability.
They involve the quantum $R$ matrices 
of the antisymmetric tensor representations of $U_p(A^{(1)}_{n-1})$
and the spin representations of 
$U_p(B^{(1)}_{n})$,
$U_p(D^{(1)}_{n})$ and $U_p(D^{(2)}_{n+1})$.

\vspace{0.4cm}

\section{Introduction}

The tetrahedron equation \cite{Zam80} is a three dimensional (3D) 
analogue of the Yang-Baxter equation \cite{Bax}.
Consider the following version of the tetrahedron equation 
often referred to as the $RLLL=LLLR$ relation:
\begin{align*}
L_{124}L_{135}L_{236}\Rm_{456} =
\Rm_{456}L_{236}L_{135}L_{124}.
\end{align*}
Here $L$ and $\Rm$ are linear operators 
on $V \otimes V \otimes F$ and $F \otimes F \otimes F$ respectively 
for some vector spaces $V$ and $F$.
The above equality is to hold between the operators on
$\overset{1}{V}\otimes
\overset{2}{V}\otimes
\overset{3}{V}\otimes
\overset{4}{F}\otimes
\overset{5}{F}\otimes
\overset{6}{F}$, where the superscripts specify the 
components on which $L$ and $\Rm$ act nontrivially.
We call the latter as 3D $\Rm$.

To grasp the structure 
let us suppress the indices for the space $F$
regarding it as {\em auxiliary} and write 
the above equation as
\begin{align*}
(L_{12}L_{13}L_{23})\Rm = \Rm(L_{23}L_{13}L_{12}).
\end{align*}
It manifests that the tetrahedron equation is 
a Yang-Baxter equation up to conjugation by the 3D $\Rm$.
One may also view it as a {\em quantized} Yang-Baxter equation
in the sense that the Boltzmann weights 
for a 2D vertex model encoded in $L$ become $\mathrm{End}(F)$ valued. 
This observation, though tautological, is known to 
lead to infinite families of 
$R$ matrices, i.e. solutions to the Yang-Baxter equation, 
in the form of {\em matrix product} \cite{S97,BS,KS}.
The method is first to form the {\em $n$-concatenation} of the original equation
by replacing the space $1$ with the copies 
$1_1,\ldots, 1_n$ and similarly for $2$ and $3$ as
\begin{align*}
(L_{1_1 2_1}L_{1_13_1}L_{2_13_1}) 
\cdots (L_{1_n 2_n }L_{1_n 3_n}L_{2_n 3_n})
\Rm=
\Rm(L_{2_13_1}L_{1_13_1}L_{1_12_1})\cdots 
(L_{2_n3_n}L_{1_n3_n}L_{1_n2_n}).
\end{align*}
One can reduce this,  after inserting spectral parameters, 
to the Yang-Baxter equation 
by evaluating $\Rm$ out suitably, for example 
by taking the trace over the auxiliary space 
$\overset{4}{F}\otimes
\overset{5}{F}\otimes
\overset{6}{F}$.
The objects that remain after the reduction 
necessarily possess the structure of $n$-matrix product over the auxiliary space.

This construction based on the 3D integrability is known to 
work efficiently for the local $L$ matrix in (\ref{air}).  
It corresponds to the choice $V = \C^2$ and 
$F$= $q$-boson Fock space $F_{q^2}$, which 
may be viewed as a $q$-boson valued six vertex model.
The resulting solutions to the Yang-Baxter equation 
are expressed in the matrix product forms (\ref{obata1}) and (\ref{obata2}).
They live in $\mathrm{End}({\bf V} \otimes {\bf V})$ with
${\bf V} = (\C^2)^{\otimes n}$, and cover the quantum $R$ matrices for the 
antisymmetric tensor representations of $U_p(A^{(1)}_{n-1})$ \cite{BS}
and the spin representations of $U_p(B^{(1)}_{n})$,
$U_p(D^{(1)}_{n})$, $U_p(D^{(2)}_{n+1})$ for some $p$ \cite{KS}.
See Appendix \ref{app:str} for the precise identification.

The purpose of this paper is to launch a similar 3D approach to the 
reflection equation \cite{Sk,Kul}. 
We propose the {\em quantized} reflection equation
\begin{align*}
(L_{12} K_2  L_{21} K_1)\Km
= \Km (K_1 L_{12} K_2  L_{21}),
\end{align*}
which is the traditional (or 2D) reflection equation up to conjugation by $\Km$.
As with the preceding illustration on the tetrahedron equation,
it actually means 
$L_{1 2 3}K_{24}L_{215}K_{16}\Km_{3456} = 
\Km_{3456}K_{16}L_{125}K_{24}L_{213}$,
where $3,4,5,6$ are labels of the auxiliary spaces suppressed in the notation.
We employ the same $L$ as before and take the local $K$ to be the 
$q$-boson valued  $2\times 2$ matrix as in (\ref{air2}).
With these choices, the quantized reflection equation may be viewed as specifying the 
{\em auxiliary linear problem} for the conjugation matrix $\Km$.
In the present setting it should be a linear operator on the Fock space 
$F_{q^2}\otimes F_{q}\otimes F_{q^2}\otimes F_{q}$. 

Our first finding is that such $\Km$ is provided 
exactly by the first nontrivial solution \cite{KO1} to the Isaev-Kulish 
3D reflection equation \cite{IK}.
We call it 3D $\Km$.
Originally the 3D $\Km$ was characterized as the intertwiner of 
the representations of the Hopf algebra 
known as the {\em quantized coordinate ring} $A_q(sp_4)$.
See Section \ref{sec:rk} for a quick exposition of the background 
\cite{D,RTF,So2} including the application to the tetrahedron equation \cite{KV}.
Our result here shows that the quantized reflection equation
and the intertwining relation for the $A_q(sp_4)$ modules in \cite{KO1}
are equivalent as auxiliary linear problems.

The reformulation of the 3D $\Km$ is quite beneficial since 
the quantized reflection equation admits, like the $RLLL=LLLR$ relation, 
the $n$-concatenation 
with respect to the auxiliary space:
\begin{align*}
(L_{1_1 2_1} K_{2_1}  L_{2_11_1} K_{1_1})
\cdots (L_{1_n 2_n} K_{2_n}  L_{2_n 1_n} K_{1_n})\Km
= \Km (K_{1_1} L_{1_1 2_1} K_{2_1}  L_{2_1 1_1}) \cdots 
(K_{1_n} L_{1_n 2_n} K_{2_n}  L_{2_n 1_n}).
\end{align*}
It is again possible to reduce this to the usual reflection equation 
by evaluating $\Km$ away 
by taking the trace or matrix elements with respect to certain eigenvectors.
These procedures are called {\em trace reduction} 
and {\em boundary vector reduction}, 
respectively\footnote{Our boundary vector reduction 
is based on the conjectural property (\ref{syki}).}.
The resulting solutions to the reflection equation  
involve the previously mentioned $R$ matrices for 
$U_p(A^{(1)}_{n-1})$, 
$U_p(B^{(1)}_{n})$,
$U_p(D^{(1)}_{n})$ and $U_p(D^{(2)}_{n+1})$. 
The companion $K$ matrices 
are trigonometric\footnote{Up to overall normalization,
matrix elements are rational 
in $q$ and the multiplicative spectral parameter $z$.} 
and are expressed in the matrix product form as in 
(\ref{sae}) and (\ref{sizka}).
They are $2^n$ by $2^n$ matrices on ${\bf V}$ which are 
neither diagonal in general nor associated with the
well-studied $R$ matrices for the 
vector representation \cite{B,Ji2}.
Therefore they are distinct from those obtained in \cite{BFKZ,ML} 
for generic $n$
and constitute new systematic solutions to the reflection equation.

The idea and maneuver in this paper demonstrate a new approach to
2D integrable systems with boundaries. 
We hope that it fuels further applications to related subjects, 
e.g. special functions, quantum many body systems, stochastic processes 
(cf. \cite{DEHP, KMO2}) and so forth in the presence of boundaries.

\vspace{0.2cm}
The layout of the paper is as follows.
In Section \ref{sec:qre} 
$q$-boson valued local $L$ and $K$ matrices are introduced and the  
quantized reflection equation is formulated.
The conjugation matrix contained therein is identified with 
the 3D $\Km$ \cite{KO1} by writing out the auxiliary linear problem explicitly.

In Section \ref{sec:rk} we
briefly review the 3D $\Rm$ and 3D $\Km$ 
based on \cite{KV,BS,KO1}.
These objects are destined to be 
eliminated in the reduction procedures in later sections.
However they essentially control the construction behind the scene in that they 
guide precisely how the local $L$ and $K$ should be combined, 
how the spectral parameters 
are to be arranged and what kind of boundary vectors are acceptable.

In Section \ref{sec:ybe} 
we recall the reduction procedures to get solutions to the Yang-Baxter equation 
from the $n$-concatenation of the $RLLL=LLLR$ relation (\ref{LLLR}).
This idea has a long history, see for example 
\cite{S97,KasV,BS,KS} and references therein.
The prescription is to eliminate the 3D $\Rm$ 
either by taking trace (trace reduction) 
or evaluating matrix elements
between certain eigenvectors of $\Rm$ (boundary vector reduction)
as already mentioned.
In our setting they reproduce
the solutions $S^{\mathrm{tr}}(z)$ \cite{BS} and 
$S^{s,s'}(z)\,(s,s'=1,2)$ \cite{KS}.
They both act on 
${\bf V} \otimes {\bf V}$ whose 
representation theoretical origin is explained in Appendix \ref{app:str}.

In Section \ref{sec:re}
we demonstrate that the same machinery works perfectly 
also for the quantized reflection equation. 
It produces the $K$ matrices 
$K^{\mathrm{tr}}(z)$ and $K^{k,k'}(z)\, (k,k'=1,2)$ that act on ${\bf V}$.
Together with the 
$S^{\mathrm{tr}}(z)$ and $S^{s,s'}(z)$ derived in Section \ref{sec:ybe},
they constitute solutions to the usual (2D) reflection equation.
This is the main result of the paper.

In Section \ref{sec:end} we give a short summary 
and mention some future problems.

Appendix \ref{app:rk} contains explicit formulas of
3D $\Rm$ and 3D $\Km$.
Appendix \ref{app:str} recalls the precise identification of the
$S^{\mathrm{tr}}(z)$ and 
$S^{s,s'}(z)$
with the quantum $R$ matrices for the 
antisymmetric tensor representations of $U_p(A^{(1)}_{n-1})$ \cite{BS}
and the spin representations of 
$U_p(D^{(2)}_{n+1})$,$U_p(B^{(1)}_{n})$, 
$U_p(D^{(1)}_{n})$ \cite{KS}, respectively.
Appendix \ref{app:ex} presents a couple of 
examples of $S^{\mathrm{tr}}(z), S^{s,s'}(z)$
and the $K$ matrices 
$K^{\mathrm{tr}}(z)$ and $K^{k,k'}(z)$.

Throughout the paper we assume that $q$ is generic and 
use the following notations:
\begin{align*}
&(z;q)_m = \prod_{k=1}^m(1-z q^{k-1}),\;\;
(q)_m = (q; q)_m,\;\;
\binom{m}{k}_{\!\!q}= \frac{(q)_m}{(q)_k(q)_{m-k}},\\
&\theta(\text{true})=1,\;\;\theta(\text{false}) = 0,\quad
{\bf e}_j = (0,\ldots,0,\overset{j}{1},0,\ldots, 0) \in \Z^n\;(1 \le j \le n).
\end{align*}

\section{Quantized reflection equation}\label{sec:qre}

\subsection{\mathversion{bold}$q$-boson valued $L$ and $K$ matrices}
Let 
$F_q = \bigoplus_{m\ge 0}\C |m\rangle$ 
and $F^\ast_q = \bigoplus_{m \ge 0} \C\langle m |$ be 
the Fock space and its dual 
equipped with the inner product 
$\langle m | m'\rangle = (q^2)_m\delta_{m,m'}$.
We define the $q$-boson operators $\ap, \am, \ok$ on them by 
\begin{equation*}
\begin{split}
&\ap |m\rangle = |m+1\rangle,\quad
\am |m\rangle = (1-q^{2m})|m-1\rangle,\quad
\ok |m\rangle = q^{m+\hf} |m\rangle,\\
&\langle m | \am = \langle m+1 |, \quad
\langle m | \ap = \langle m-1| (1-q^{2m}),\quad
\langle m | \ok = \langle m| q^{m+\hf}.
\end{split}
\end{equation*}
They satisfy $(\langle m | X)|m'\rangle 
= \langle m | (X|m'\rangle)$ and 
\begin{align}\label{misaki40}
&\langle m |X_1\cdots X_j |m'\rangle = 
\langle m' |\overline{X_j}\cdots \overline{X_1} |m\rangle,
\end{align}
where $\overline{(\cdots)}$ is defined by 
$\overline{\apm} = \amp,\quad \overline{\ok}=\ok$.

Let $F_{q^2}, F^\ast_{q^2}$ 
and $\Ap, \Am, \OK$ denote 
the same objects with $q$ replaced by $q^2$, i.e.,       
\begin{equation*}
\begin{split}
&\Ap |m\rangle = |m+1\rangle,\quad
\Am |m\rangle = (1-q^{4m})|m-1\rangle,\quad
\OK |m\rangle = q^{2m+1} |m\rangle,\\
&\langle m | \Am = \langle m+1 |, \quad
\langle m | \Ap = \langle m-1| (1-q^{4m}),\quad
\langle m | \OK = \langle m| q^{2m+1}.
\end{split}
\end{equation*}
The inner product in $F_{q^2}$ is given by 
$\langle m | m'\rangle = (q^4)_m\delta_{m,m'}$
differing from the $F_q$ case. 
However we write the base vectors as $\langle m |, |m\rangle$ 
either for 
$F^\ast_{q^2}, F_{q^2}$ or $F^\ast_{q}, F_{q}$
since their distinction will always be evident from the context.
Note the $q$-boson commutation relations
\begin{align}
&\ok \,\apm = q^{\pm 1}\apm \ok,\quad
\apm \amp = 1 - q^{\mp 1}\ok^2,
\label{ngh1}\\
&\OK\, \Apm = q^{\pm 2}\Apm \OK,\quad
\Apm \Amp = 1 - q^{\mp 2}\OK^2.
\label{ngh2}
\end{align}
We will also use the number operator ${\bf h}$ defined by 
\begin{align}\label{syri}
{\bf h}|m\rangle = m | m\rangle,\qquad
\langle m | {\bf h}= \langle m| m
\end{align}
either for $F_q$ or $F_{q^2}$.
One may regard $\ok=q^{{\bf h}+\hf}$ and 
$\OK = q^{2{\bf h}+1}$.
The extra $1/2$ in the spectrum of $\log_q \ok$
is the celebrated {\em zero point energy}.
It makes the forthcoming equations
(\ref{yuna1})--(\ref{yuna2}) and (\ref{recR}) 
totally free from the apparent $q$. (This is an 
indication of a parallel story in the modular double setting.)

Set $V = \C v_0 \oplus \C v_1\simeq \C^2$ and 
introduce the $q$-boson valued 
$L$ matrices and $K$ matrices by
\begin{align}\label{air}
&L= \begin{pmatrix}
L_{0,0}^{0,0} & 
L_{0,1}^{0,0} & 
L_{1,0}^{0,0} & 
L_{1,1}^{0,0} \\
L_{0,0}^{0,1} & 
L_{0,1}^{0,1} & 
L_{1,0}^{0,1} & 
L_{1,1}^{0,1} \\
L_{0,0}^{1,0} & 
L_{0,1}^{1,0} & 
L_{1,0}^{1,0} & 
L_{1,1}^{1,0} \\
L_{0,0}^{1,1} & 
L_{0,1}^{1,1} & 
L_{1,0}^{1,1} & 
L_{1,1}^{1,1} 
\end{pmatrix} 
= \begin{pmatrix}
1 & 0 & 0 & 0 \\
0 & \OK & \Am & 0\\
0 & \Ap & -\OK & 0\\
0 & 0 & 0 & 1
\end{pmatrix} \in \mathrm{End}(V \otimes V \otimes F_{q^2}),\\
&K = \begin{pmatrix}
K_{0}^{0} & 
K_{1}^{0}\\
K_{0}^{1} & 
K_{1}^{1}
\end{pmatrix}
= \begin{pmatrix}
\ap & -\ok \\
\ok & \am
\end{pmatrix} \in \mathrm{End}(V \otimes F_q).
\label{air2}
\end{align}
We let them act on the base vectors by the following rule:
\begin{align*}
L(v_\alpha \otimes v_\beta \otimes |m\rangle)
&= \sum_{0 \le \gamma,\delta \le 1}v_\gamma\otimes v_\delta \otimes 
L^{\gamma, \delta}_{\alpha, \beta}|m\rangle,
\quad
K(v_\alpha \otimes |m\rangle)
= \sum_{0 \le \beta \le 1}v_\beta\otimes 
K^{\beta}_{\alpha}|m\rangle.
\end{align*}
Explicitly they read
\begin{align*}
L(v_0\otimes v_0 \otimes |m\rangle)
&= v_0 \otimes v_0 \otimes |m\rangle,
\quad
L(v_1\otimes v_1 \otimes |m\rangle)
= v_1 \otimes v_1 \otimes |m\rangle,
\\
L(v_0\otimes v_1 \otimes |m\rangle)
&= v_0 \otimes v_1 \otimes \OK|m\rangle
+ v_1 \otimes v_0 \otimes \Ap|m\rangle
=  q^{2m+1}v_0 \otimes v_1 \otimes |m\rangle
+ v_1 \otimes v_0 \otimes |m+1\rangle,
\\
L(v_1\otimes v_0 \otimes |m\rangle)
&= v_0 \otimes v_1 \otimes \Am|m\rangle
- v_1 \otimes v_0 \otimes \OK|m\rangle
= (1-q^{4m})v_0 \otimes v_1 \otimes |m-1\rangle
- q^{2m+1}v_1 \otimes v_0 \otimes |m\rangle,
\\
K(v_0 \otimes |m \rangle) &= v_0 \otimes \ap |m\rangle 
+ v_1 \otimes \ok|m\rangle
=v_0 \otimes |m+1\rangle 
+ q^{m+\hf}v_1 \otimes |m\rangle,
\\
K(v_1 \otimes |m \rangle) &= -v_0 \otimes \ok |m\rangle 
+ v_1 \otimes \am |m\rangle
=-q^{m+\hf}v_0 \otimes |m\rangle 
+ (1-q^{2m}) v_1 \otimes |m-1\rangle.
\end{align*}
Note the obvious properties
\begin{align}
&L^{\gamma,\delta}_{\alpha,\beta}
=0
\quad \text{unless}\;\; \alpha+\beta=\gamma+\delta,
\label{mzsma1}\\
&{\bf h} L^{\gamma,\delta}_{\alpha,\beta}
=L^{\gamma,\delta}_{\alpha,\beta}
({\bf h}+\beta-\delta),\qquad
{\bf h}K^{\beta}_{\alpha}
= K^{\beta}_{\alpha}({\bf h}+1-\alpha-\beta),
\label{mzsma2}
\end{align}
which will be referred to as {\em weight conservation}.
We depict $L$ as
\begin{align*}
\begin{picture}(350.5,72)(6,35)
\put(12,80){
\put(-11,0){\vector(1,0){22}}\put(0,-10){\vector(0,1){20}}
}
\multiput(80,80)(55,0){6}{
\put(-11,0){\vector(1,0){22}}\put(0,-10){\vector(0,1){20}}
}
\put(-68,0){
\put(60,77){$\alpha$}\put(77.5,60){$\beta$}
\put(94.5,77){$\gamma$}\put(77.5,94){$\delta$}
}
\put(61,77){0}\put(77.5,60){0}\put(94,77){0}\put(77.5,94){0}
\put(55,0){
\put(61,77){1}\put(77.5,60){1}\put(94,77){1}\put(77.5,94){1}
}
\put(110,0){
\put(61,77){0}\put(77.5,60){1}\put(94,77){0}\put(77.5,94){1}
}
\put(165,0){
\put(61,77){1}\put(77.5,60){0}\put(94,77){1}\put(77.5,94){0}
}
\put(220,0){
\put(61,77){0}\put(77.5,60){1}\put(94,77){1}\put(77.5,94){0}
}
\put(275,0){
\put(61,77){1}\put(77.5,60){0}\put(94,77){0}\put(77.5,94){1}
}
\put(78,40){
\put(-74,0){$L^{\gamma,\delta}_{\alpha,\beta}$}
\put(0,0){1} \put(55,0){1} \put(109,0){$\OK$} 
\put(157,0){$-\OK$} \put(218,0){$\Ap$} \put(274,0){$\Am$}
}
\end{picture}
\end{align*}
So $L$ may be regarded as defining a $q$-boson valued six vertex model
in which the latter relation of (\ref{ngh2}) plays the role of  
``free-fermion" condition.
See eq. $(10.16.5)|_{d=0}$ in \cite{Bax}.
Similarly $K$ is pictured as
\begin{align}\label{kfig}
\begin{picture}(200,50)(-55,-30)
\put(-70,0){
\put(0,-12){\line(0,1){24}}\put(0,0){\line(-1,-1){10}}
\put(0,0){\vector(-1,1){10}}
}
\put(-80,-30){$K^{\beta}_{\alpha}$}
\put(-71,0){\put(-17,-16){$\alpha$}\put(-17.5,9){$\beta$}}
\multiput(0,0)(55,0){4}{
\put(0,-12){\line(0,1){24}}\put(0,0){\line(-1,-1){10}}
\put(0,0){\vector(-1,1){10}}
}
\put(-17,-16){0}\put(-17.5,9){0}
\put(55,0){\put(-17,-16){0}\put(-17.5,9){1}}
\put(110,0){\put(-17,-16){1}\put(-17.5,9){0}}
\put(165,0){\put(-17,-16){1}\put(-17.5,9){1}}
\put(-9,-31){$\ap$}
\put(49,-31){$\ok$}
\put(98,-31){$-\ok$}
\put(157,-31){$\am$}
\end{picture}
\end{align} 
Here the lines without an arrow signifies a reflecting boundary
to which no physical degree of freedom is assigned.
One may imagine that each vertex is a $q$-boson operator
acting in the direction perpendicular to these planar diagrams
from the back to the front. 
Up to conventional difference, the  
$q$-boson valued $L$ matrix (\ref{air}) 
appeared in \cite{BS}.

\subsection{Quantized reflection equation}
Let 
$\Km \in \mathrm{End}(F_{q^2} \otimes F_q \otimes F_{q^2} \otimes F_q)$
be a linear operator so normalized as
\begin{align}\label{tsgmi}
\Km (|0 \rangle \otimes |0 \rangle \otimes
|0 \rangle \otimes|0 \rangle) = 
|0 \rangle \otimes |0 \rangle \otimes
|0 \rangle \otimes|0 \rangle.
\end{align}
For $L$ and $K$ defined in (\ref{air}) and (\ref{air2}),
we propose the $q$-boson valued reflection equation
that holds up to $\Km$-conjugation as follows:
\begin{align}\label{grk}
(L_{12} K_2  L_{21} K_1)\Km
= \Km (K_1 L_{12} K_2  L_{21}).
\end{align}
This is an equality of linear operators on
$\overset{1}{V}\otimes 
\overset{2}{V} \otimes \overset{3}{F}_{q^2} 
\otimes \overset{4}{F}_q \otimes \overset{5}{F}_{q^2} 
\otimes \overset{6}{F}_q$,
where the superscripts are just temporal labels for explanation. 
If they are all exhibited (\ref{grk}) reads as 
\begin{align}\label{hrk}
L_{1 2 3}K_{24}L_{215}K_{16}\Km_{3456} = 
\Km_{3456}K_{16}L_{125}K_{24}L_{213}.
\end{align}
Here $L_{125}$ for example stands for the operator that acts as $L$ on 
$\overset{1}{V}\otimes \overset{2}{V} \otimes \overset{5}{F}_{q^2}$
and as the identity elsewhere.
We have also set 
$L_{215} = P_{1 2} L_{125} P_{12}$ and 
$L_{213} = P_{1 2} L_{123} P_{12}$,
where $P_{12}: u \otimes v \mapsto v \otimes u$ is the 
interchange of the components of 
$\overset{1}{V}\otimes \overset{2}{V}$.
Put in words, the factors 
$L_{12}K_2  L_{21} K_1$ 
and $K_1  L_{12}  K_2  L_{21}$ 
in (\ref{grk}) mean the compositions of operators on 
$\overset{1}{V}\otimes \overset{2}{V}$
and simultaneously represent 
the tensor product corresponding to the Fock part 
$\overset{3}{F}_{q^2} 
\otimes \overset{4}{F}_q \otimes \overset{5}{F}_{q^2} 
\otimes \overset{6}{F}_q$.
The multiplication of $\Km$ is with respect to the Fock part.
When $\Km$ is trivial and $L$ and $K$ become scalars on 
$\overset{3}{F}_{q^2} 
\otimes \overset{4}{F}_q \otimes \overset{5}{F}_{q^2} 
\otimes \overset{6}{F}_q$, the equation (\ref{grk}) 
reduces to the usual reflection equation without a spectral parameter
$L_{12}K_2L_{21}K_1 = K_1L_{12}K_2L_{21}$.
In this sense we call (\ref{grk}) 
the {\em quantized reflection equation}.
Schematically the equation (\ref{hrk}) is shown as follows:
\begin{align*}
\begin{picture}(200,94)(-20,-15)
\put(0,-9){\line(0,1){90}}
\put(0,20){\line(-1,-2){13}}\put(0,20){\vector(-1,2){30}}
\put(0,50){\line(-2,-1){40}}\put(0,50){\vector(-2,1){40}}
\put(-18,63){3}\put(4,46){4}\put(-16.7,33){5}\put(4,16){6}
\put(-48,25){2} \put(-19,-14){1}
\put(20,30){$\circ\; \,\Km_{3456}\;\;=\;\; \Km_{3456}\,\;\circ$}
\put(175,0){
\put(0,-9){\line(0,1){90}}
\put(0,50){\line(-1,-2){28}}\put(0,50){\vector(-1,2){15}}
\put(0,20){\line(-2,-1){40}}\put(0,20){\vector(-2,1){40}}
\put(-18.3,1.2){3}\put(4,16){4}\put(-17,32){5}\put(4,46){6}
\put(-48,-4){2} \put(-34,-14){1}
}
\end{picture}
\end{align*}
Here the indices $1,2$ are assigned to the lines, whereas
$3,4,5,6$ are attached to the vertices.
One may rather regard $\Km_{3456}$ in the left (right) hand side
as a point in the back (front) of the diagram where 
the four arrows going toward (coming from) 
the vertices $3,4,5,6$ intersect\footnote{
We did not have the graphical skill to draw such a nice figure.}.

Let us write down (\ref{grk}) explicitly.
Let $(abij)$ denote its matrix element 
corresponding to the transition 
$v_j\otimes v_i \mapsto v_b \otimes v_a$ in 
$\overset{1}{V} \otimes \overset{2}{V}$.
(The strange inversion of the indices is just by a conventional reason to take 
contact with \cite{KO1}.)
They read as 
\begin{alignat}{2}
&( 1111 ): & \quad &
[ \ichi \!\otimes\! \am \!\otimes\! \ichi \!\otimes\! \am
-  \ichi \!\otimes\! \ok \!\otimes\! \Am \!\otimes\! \ok, \,\Km] = 0,
\label{yuna1}\\
&( 1110 ): & \quad &
(\ichi \!\otimes\! \am \!\otimes\! \ichi \!\otimes\! \ok
+ \ichi \!\otimes\! \ok \!\otimes\! \Am \!\otimes\! \ap)\Km \nonumber\\
&&& =
\Km(\Am \!\otimes\! \ap \!\otimes\! \Am \!\otimes\! \ok+
\Am \!\otimes\! \ok \!\otimes\! \ichi \!\otimes\! \am -
\OK \!\otimes\! \am \!\otimes\! \OK \!\otimes\! \ok),\\
&( 1101 ): & \quad &
(\ichi \!\otimes\! \ok \!\otimes\! \OK \!\otimes\! \am) \Km = 
\Km(
\Ap \!\otimes\! \am \!\otimes\! \OK \!\otimes\! \ok + 
\OK \!\otimes\! \ap \!\otimes\! \Am \!\otimes\! \ok +
\OK \!\otimes\! \ok \!\otimes\! \ichi \!\otimes\! \am),\\
&( 1100 ): & \quad &
[\ichi \!\otimes\! \ok \!\otimes\! \OK \!\otimes\! \ok, \Km] = 0,\\
&( 1011 ): & \quad &
(\Am \!\otimes\! \ap \!\otimes\! \Am \!\otimes\! \ok
+\Am \!\otimes\! \ok \!\otimes\! \ichi \!\otimes\! \am -
\OK \!\otimes\! \am \!\otimes\! \OK \!\otimes\! \ok)\Km \nonumber\\
&&& =\Km( \ichi \!\otimes\! \am \!\otimes\! \ichi \!\otimes\! \ok +
\ichi \!\otimes\! \ok \!\otimes\! \Am \!\otimes\! \ap),\\
&( 1010 ): & \quad &
[\Am \!\otimes\! \ap \!\otimes\! \Am \!\otimes\! \ap -
\Am \!\otimes\! \ok \!\otimes\! \ichi \!\otimes\! \ok -
\OK \!\otimes\! \am \!\otimes\! \OK \!\otimes\! \ap, \,\Km] = 0,\\
&( 1001 ): & \quad &
(\Am \!\otimes\! \ap \!\otimes\! \OK \!\otimes\! \am + 
\OK \!\otimes\! \am \!\otimes\! \Ap \!\otimes\! \am - 
\OK \!\otimes\! \ok \!\otimes\! \ichi \!\otimes\! \ok)\Km \nonumber\\
&&&=
\Km(\Ap \!\otimes\! \am \!\otimes\! \OK \!\otimes\! \ap+
\OK \!\otimes\! \ap \!\otimes\! \Am \!\otimes\! \ap -
\OK \!\otimes\! \ok \!\otimes\! \ichi \!\otimes\! \ok),\\
&( 1000 ): & \quad &
(\Am \!\otimes\! \ap \!\otimes\! \OK \!\otimes\! \ok+
\OK \!\otimes\! \am \!\otimes\! \Ap \!\otimes\! \ok+
\OK \!\otimes\! \ok \!\otimes\! \ichi \!\otimes\! \ap )\Km=
\Km(
\ichi \!\otimes\! \ok \!\otimes\! \OK \!\otimes\! \ap),\\
&( 0111 ): & \quad &
(\Ap \!\otimes\!\am \!\otimes\! \OK \!\otimes\! \ok + 
\OK \!\otimes\! \ap \!\otimes\! \Am \!\otimes\! \ok +
\OK \!\otimes\! \ok \!\otimes\! \ichi \!\otimes\! \am)\Km
=
\Km(\ichi \!\otimes\! \ok \!\otimes\! \OK \!\otimes\! \am),\\
&( 0110 ): & \quad &
(\Ap \!\otimes\! \am \!\otimes\! \OK \!\otimes\! \ap+
\OK \!\otimes\! \ap \!\otimes\! \Am \!\otimes\! \ap - 
\OK \!\otimes\! \ok \!\otimes\! \ichi \!\otimes\! \ok)\Km \nonumber\\
&&& =
\Km(\Am \!\otimes\! \ap \!\otimes\! \OK \!\otimes\! \am +
\OK \!\otimes\! \am \!\otimes\! \Ap \!\otimes\! \am -
\OK \!\otimes\! \ok \!\otimes\! \ichi \!\otimes\! \ok),\\
&( 0101 ): & \quad &
[\Ap \!\otimes\! \am \!\otimes\! \Ap \!\otimes\! \am - 
\Ap \!\otimes\! \ok \!\otimes\! \ichi \!\otimes\! \ok -
\OK \!\otimes\! \ap \!\otimes\! \OK \!\otimes\! \am,\, \Km]=0,\\
&( 0100 ): & \quad &
(\Ap \!\otimes\! \am \!\otimes\! \Ap \!\otimes\! \ok +
\Ap \!\otimes\! \ok \!\otimes\! \ichi \!\otimes\! \ap -
\OK \!\otimes\! \ap \!\otimes\! \OK \!\otimes\! \ok)\Km \nonumber\\
&&&= \Km(\ichi \!\otimes\! \ap \!\otimes\! \ichi \!\otimes\! \ok +
\ichi \!\otimes\! \ok \!\otimes\! \Ap \!\otimes\! \am),\\
&( 0011 ): & \quad &
[\ichi \!\otimes\! \ok \!\otimes\! \OK \!\otimes\! \ok,\, \Km] = 0
\quad 
(\text{same as $( 1100 )$}),\\
&( 0010 ): & \quad &
(\ichi \!\otimes\! \ok \!\otimes\! \OK \!\otimes\! \ap)\Km = 
\Km(\Am \!\otimes\! \ap \!\otimes\! \OK \!\otimes\! \ok +
\OK \!\otimes\! \am \!\otimes\! \Ap \!\otimes\! \ok+
\OK \!\otimes\! \ok \!\otimes\! \ichi \!\otimes\! \ap),\\
&( 0001 ): & \quad &
(\ichi \!\otimes\! \ap \!\otimes\! \ichi \!\otimes\! \ok + 
\ichi \!\otimes\! \ok \!\otimes\! \Ap \!\otimes\! \am)\Km \nonumber\\
&&&=\Km(\Ap \!\otimes\! \am \!\otimes\! \Ap \!\otimes\! \ok+
\Ap \!\otimes\! \ok \!\otimes\! \ichi \!\otimes\! \ap - 
\OK \!\otimes\! \ap \!\otimes\! \OK \!\otimes\! \ok),\\
&( 0000 ): & \quad &
[ \ichi \!\otimes\! \ap \!\otimes\! \ichi \!\otimes\! \ap -
\ichi \!\otimes\! \ok \!\otimes\! \Ap \!\otimes\! \ok,\, \Km] = 0.
\label{yuna2}
\end{alignat}
As an illustration the second equation $(1110)$ originates in the 
matrix element for the transition $v_0 \otimes v_1 \mapsto v_1\otimes v_1$.
The corresponding factors 
$L_{1 2 3}K_{24}L_{215}K_{16}$ and 
$K_{16}L_{125}K_{24}L_{213}$
in (\ref{hrk}) are calculated as
\begin{align*}
\begin{picture}(400,130)(-50,-36)
\put(0,-9){\line(0,1){90}}
\put(0,20){\line(-1,-2){13}}\put(0,20){\vector(-1,2){30}}
\put(0,50){\line(-2,-1){40}}\put(0,50){\vector(-2,1){40}}
\put(-47,70){$\si$}\put(-22,48){$\si$}
\put(-34,82){$\si$}\put(-11,59){$\si$}\put(-7,40.5){$\si$}
\put(-12,27){$\si$}
\put(-47,27){$\si$} \put(-17,-13){$\sz$}
\put(-50,-30){$1\!\otimes\!\am \!\otimes\!1 \!\otimes\! \ok$}
\put(20,-30){$1\!\otimes\! \ok\!\otimes\! \Am \!\otimes\! \ap$}
\put(15,35){$+$}
\put(80,0){
\put(0,-9){\line(0,1){90}}
\put(0,20){\line(-1,-2){13}}\put(0,20){\vector(-1,2){30}}
\put(0,50){\line(-2,-1){40}}\put(0,50){\vector(-2,1){40}}
\put(-47,70){$\si$}\put(-22,48){$\si$}
\put(-34,82){$\si$}\put(-11,59){$\si$}\put(-7,40.5){$\sz$}
\put(-12,27){$\sz$}
\put(-47,27){$\si$} \put(-17,-13){$\sz$}
}
\put(170,0){
\put(15,35){$+$}\put(95,35){$+$}
\put(0,-9){\line(0,1){90}}
\put(0,50){\line(-1,-2){28}}\put(0,50){\vector(-1,2){15}}
\put(0,20){\line(-2,-1){40}}\put(0,20){\vector(-2,1){40}}
\put(-46,40){$\si$}\put(-11,8){$\sz$}\put(-22,18){$\si$}
\put(-19,83){$\si$}\put(-11,40.5){$\sz$}\put(-7.5,25.5){$\sz$}
\put(-47,-3){$\si$} \put(-31,-13){$\sz$}
\put(-60,-30){$\Am\!\otimes\! \ap \!\otimes\! \Am \!\otimes\! \ok$}
\put(80,0){
\put(0,-9){\line(0,1){90}}
\put(0,50){\line(-1,-2){28}}\put(0,50){\vector(-1,2){15}}
\put(0,20){\line(-2,-1){40}}\put(0,20){\vector(-2,1){40}}
\put(-46,40){$\si$}\put(-11,8){$\sz$}\put(-22,18){$\si$}
\put(-19,83){$\si$}\put(-11,40.5){$\si$}\put(-7.5,25.5){$\si$}
\put(-47,-3){$\si$} \put(-31,-13){$\sz$}
\put(-50,-30){$\Am\!\otimes\! \ok \!\otimes\! 1 \!\otimes\! \am$}
}
\put(160,0){
\put(0,-9){\line(0,1){90}}
\put(0,50){\line(-1,-2){28}}\put(0,50){\vector(-1,2){15}}
\put(0,20){\line(-2,-1){40}}\put(0,20){\vector(-2,1){40}}
\put(-46,40){$\si$}\put(-11,8){$\si$}\put(-22,18){$\sz$}
\put(-19,83){$\si$}\put(-11,40.5){$\sz$}\put(-7.5,25.5){$\si$}
\put(-47,-3){$\si$} \put(-31,-13){$\sz$}
\put(-50,-30){$-\OK\!\otimes\! \am \!\otimes\! \OK \!\otimes\! \ok$}
}
}
\end{picture}
\end{align*}

One of our main findings in this paper is that 
the quantized reflection equation (\ref{grk}) 
listed in (\ref{yuna1})--(\ref{yuna2}) exactly reproduces the 
characterization condition (\ref{reK}) of the 3D reflection matrix 
as the intertwiner of the quantized coordinate ring 
$A_q(sp_4)$ in \cite[App.A]{KO1}\footnote{The $q$-boson operators 
$\OK$ and $\ok$ in \cite{KO1} do not contain the zero point energy, hence
there are powers of $q$ around.}.
Therefore $\Km$ {\em is} the 3D reflection matrix.
We will make use of this connection and consequent properties efficiently 
in Section \ref{sec:ybe} and Section \ref{sec:re}.
The next section is meant to be a preparation for it recalling 
basic facts from \cite{KV,BS,KO1}.
In short we have found a solution to the quantized reflection equation.

\vspace{0.2cm}
{\em Remark}. The set of equations (\ref{yuna1})--(\ref{yuna2}) is 
invariant under the exchange 
$\ap \leftrightarrow \am,
\Ap \leftrightarrow \Am$.
Therefore  
$(L',K') = (L,K)|_{\ap \leftrightarrow \am,
\Ap \leftrightarrow \Am}$ can also be employed 
to formulate the quantized reflection equation to characterize the same $\Km$.
The reduction procedure in Section \ref{sec:ybe} and \ref{sec:re} works
even for the {\em mixture} of $L, L',K, K'$.
However the resulting degree of  freedom which apparently 
generalizes (\ref{obata1}), (\ref{obata2}), 
(\ref{sae}) and (\ref{sizka}) 
can be absorbed into a suitable redefinition of the bases and gauges.

\section{Brief summary on 3D $R$ and 3D $K$}\label{sec:rk}
\subsection{\mathversion{bold}
$A_q(sl_3)$, 3D $\Rm$ and tetrahedron equation}
The quantized coordinate ring $A_q(sl_3)$ is a Hopf algebra
realized by 9 generators $(t_{ij})_{1 \le i,j \le 3}$ with relations \cite{RTF}.
Their explicit form is available in \cite[Sec.2]{KO1}.
The maps
\begin{align}\label{mih0}
\pi_1: (t_{ij})_{1 \le i,j \le 3} \mapsto 
\begin{pmatrix}
\am & c_1\ok & 0\\
d_1\ok& \ap  & 0\\
0 & 0 & 1 \\
\end{pmatrix},\;
\;\;
\pi_2: (t_{ij})_{1 \le i,j \le 3} \mapsto 
\begin{pmatrix}
1 & 0 & 0 \\
0 & \am &  c_2 \ok\\
0 & d_2\ok & \ap\\
\end{pmatrix}
\end{align}
with $c_1d_1=c_2d_2=-1$
give irreducible representations
$\pi_i: A_q(sl_3) \rightarrow \mathrm{End}(F_q)$.
Let us denote $\pi_i \otimes \pi_j \otimes \pi_i$ by $\pi_{iji}$ 
for short.
According to the general theory \cite{So2} (see \cite[Th.2.2]{KO1}),
the tensor product representations 
$\pi_{121} \circ \Delta$ and $\pi_{212} \circ \Delta:
A_q(sl_3) \rightarrow \mathrm{End}((F_q)^{\otimes 3})$
are both irreducible and equivalent.
Here $\Delta: A_q(sl_3) \rightarrow A_q(sl_3)^{\otimes 3}$ 
is the coproduct specified for the generators as
$\Delta(t_{ij}) = \sum_{l_1 l_2}t_{i l_1}\otimes t_{l_1 l_2} \otimes t_{l_2 j}$.
Therefore there is a unique map
$\Phi: (F_q)^{\otimes 3} \rightarrow (F_q)^{\otimes 3}$ characterized by
the intertwining relation and the normalization as follows \cite{KV}:
\begin{align}
&\pi_{212}(\Delta(g))\circ \Phi = \Phi \circ \pi_{121}(\Delta(g))
\quad (\forall g \in A_q(sl_3)),
\label{mih1}\\
&\Phi(|0 \rangle \otimes |0 \rangle \otimes |0 \rangle)
= |0 \rangle \otimes |0 \rangle \otimes |0 \rangle.
\label{mih2}
\end{align}
We define $\hat{\Rm}$ and $\Rm$ as
\begin{align*}
\hat{\Rm}= \Phi P_{13}: (F_q)^{\otimes 3} \rightarrow (F_q)^{\otimes 3},
\qquad \Rm = \hat{\Rm}|_{q \rightarrow q^2}:
(F_{q^2})^{\otimes 3} \rightarrow (F_{q^2})^{\otimes 3},
\end{align*} 
where $P_{13}: x_1 \otimes x_2 \otimes x_3 \mapsto 
x_3 \otimes x_2 \otimes x_1$ is the linear operator 
reversing the order of the tensor product.
The conditions (\ref{mih1}) and (\ref{mih2}) are cast into 
\begin{align}
&\pi_{212}(\Delta(g))\circ \hat{\Rm} = 
\hat{\Rm} \circ \pi_{121}(\Delta'(g))\quad
(\forall g \in A_q(sl_3)),
\label{sakra1}\\
&\hat{\Rm}(|0 \rangle \otimes |0 \rangle \otimes |0 \rangle)
= |0 \rangle \otimes |0 \rangle \otimes |0 \rangle,
\label{sakra2}
\end{align}
where $\Delta'= P_{13}\Delta P_{13}$ is the opposite coproduct.
The equations (\ref{sakra1}) are actually  
independent of the parameters $c_i,d_i$ in (\ref{mih0}) as long as
$c_id_i=-1$.
We present them and 
an explicit formula of $\hat{\Rm}$ in Appendix \ref{app:rk}. 

The intertwining relation (\ref{sakra1}) admits an alternative formulation 
in a spirit closer to ``$RLL=LLR$" \cite{BS}. 
In fact in terms of $L$ in (\ref{air})  
the set of equations (\ref{sakra1}) turn out to be equivalent with 
the tetrahedron equation of $RLLL=LLLR$ type mentioned in the introduction:
\begin{align}
&L_{124}L_{135} L_{236} \,\Rm_{456} = 
\Rm_{456}L_{236}L_{135}L_{124} 
\in \mathrm{End}(
\overset{1}{V}\otimes 
\overset{2}{V}\otimes 
\overset{3}{V}  
\otimes \overset{4}{F}_{q^2} \otimes \overset{5}{F}_{q^2} 
\otimes \overset{6}{F}_{q^2}),
\label{LLLR}
\end{align}
where the notation is similar to (\ref{hrk}).

One has the symmetry $\Rm_{123} = \Rm_{321} (:=P_{13}R_{123}P_{31})$.
Some other notable properties of $\Rm$ are
\begin{align}
\text{tetrahedron eq.}\; : \;\;
&\Rm_{124} \Rm_{135} \Rm_{236} \Rm_{456} = 
\Rm_{456} \Rm_{236} \Rm_{135} \Rm_{124},
\label{ykn1}\\
\text{inversion relation}\; : \;\;
&\Rm = \Rm^{-1},
\label{ykn2}\\
\text{weight conservation}\; : \;\;
& [x^{{\bf h}_1}(xy)^{{\bf h}_2} y^{{\bf h}_3}, \Rm]=0,
\label{ykn3}
\end{align}
where $x,y$ are generic parameters and 
${\bf h}_1= {\bf h} \otimes 1 \otimes 1$,
${\bf h}_2= 1 \otimes {\bf h} \otimes 1$,
${\bf h}_3=  1 \otimes 1 \otimes {\bf h} $ in terms of 
${\bf h}$ defined in (\ref{syri}).
The weight conservation (\ref{ykn3})
will be the source of introducing {\em spectral parameters} in the reduction 
procedure.
It originates in the factor $\delta^{a+b}_{i+j}\delta^{b+c}_{j+k}$ in (\ref{Rex}).

The solution $\hat{\Rm}$ to the tetrahedron equation was obtained in 
\cite{KV}\footnote{
The formula for it on p194 in \cite{KV} contains a misprint unfortunately.
Eq. (\ref{Rex}) here is a correction of it.}
based on the representation theory of 
the quantized coordinate ring \cite{So2}.
Later it was also found from (\ref{LLLR})  
via a quantum geometry consideration \cite{BS}.
The two $\Rm$'s were identified in \cite[eq.(2.27)]{KO1}.
Here we simply call it the 3D $\Rm$.

\subsection{\mathversion{bold}$A_q(sp_4)$, 3D $\Km$ and 3D reflection equation}
The quantized coordinate ring $A_q(sp_4)$ is a Hopf algebra
realized by 16 generators $(t_{ij})_{1 \le i,j \le 4}$ with relations \cite{RTF}.  
Their explicit form is available in \cite[Sec.3]{KO1}.
The maps
\begin{align}\label{mih4}
\pi_1: (t_{ij})_{1 \le i,j \le 4} \mapsto 
\begin{pmatrix}
\am & c_1\ok & 0& 0\\
d_1\ok& \ap  & 0& 0\\
0 & 0 & \am & d_1^{-1}\ok \\
0 & 0 & c_1^{-1}\ok & \ap
\end{pmatrix},\;
\;\;
\pi_2: (t_{ij})_{1 \le i,j \le 4} \mapsto 
\begin{pmatrix}
1 & 0 & 0 & 0\\
0 & \Am & c_2 \OK & 0\\
0 & d_2 \OK & \Ap & 0\\
0 & 0 & 0  & 1\\
\end{pmatrix}
\end{align}
with $c_1 d_1 = c_2 d_2=-1$ 
give irreducible representations 
$\pi_i: A_q(sp_4) \rightarrow \mathrm{End}(F_{q^i})$ \cite{KO1}.
Coexistence of the $q$-boson and the $q^2$-boson originates in the
two distinct length of the simple roots of $sp_4$.
Let us denote $\pi_i \otimes \pi_j \otimes \pi_i \otimes \pi_j$ by 
$\pi_{ijij}$ for short.
According to the general theory \cite{So2}, 
the tensor product representations
$\pi_{1212}\circ \Delta: A_q(sp_4) \rightarrow 
\mathrm{End}(F_q\otimes F_{q^2}\otimes F_q\otimes F_{q^2})$ 
and 
$\pi_{2121}\circ \Delta: A_q(sp_4) \rightarrow 
\mathrm{End}(F_{q^2}\otimes F_q\otimes F_{q^2} \otimes F_q)$ 
are both irreducible and equivalent.
Here $\Delta: A_q(sp_4) \rightarrow A_q(sp_4)^{\otimes 4}$ 
is the coproduct specified for generators as
$\Delta(t_{ij}) = \sum_{l_1,l_2,l_3} 
t_{i l_1}\otimes t_{l_1 l_2}\otimes t_{l_2 l_3}\otimes t_{l_3 j}$.
Therefore there is a unique map
$
\Psi : 
F_{q} \otimes F_{q^2}\otimes 
F_{q}\otimes F_{q^2} \longrightarrow
F_{q^2} \otimes F_{q}
\otimes F_{q^2}\otimes F_{q}
$
characterized by the intertwining relation and the normalization:
\begin{align}
&\pi_{2121}(\Delta(g))\circ \Psi = \Psi \circ \pi_{1212}(\Delta(g))
\quad (\forall g \in A_q(sp_4)),\label{pip}\\
&\Psi (|0\rangle \otimes|0\rangle \otimes|0\rangle \otimes|0\rangle) 
=|0\rangle \otimes|0\rangle \otimes|0\rangle \otimes|0\rangle. 
\label{pin}
\end{align}
We define 3D $\Km$ by
\begin{align}\label{kyuki}
\mathscr{K} = \Psi  P_{1234} : 
\; F_{q^2} \otimes F_{q}\otimes 
F_{q^2}\otimes F_{q}
\longrightarrow 
F_{q^2} \otimes F_{q}\otimes 
F_{q^2}\otimes F_{q},
\end{align}
where 
$P_{1234} : x_1 \otimes x_2 \otimes x_3 \otimes x_4 \mapsto
x_4 \otimes x_3 \otimes x_2 \otimes x_1$ is the linear operator 
reversing the order of the tensor product.
The condition (\ref{pip}) is cast into 
\begin{align}
\pi_{2121}(\Delta(g))\circ \mathscr{K} 
= \mathscr{K} \circ \pi_{2121}(\Delta'(g))
\quad (\forall g \in A_q(sp_4)),\label{reK}
\end{align}
where $\Delta'=P_{1234}\Delta P_{1234}$ is the opposite coproduct.
With the choice $g= t_{ij}\, (1 \le i,j \le 4)$, 
(\ref{reK}) produces 16 equations.
They are independent of the parameters $c_i, d_i$ in (\ref{mih4})
as long as $c_id_i=-1$.
As stated in the end of Section \ref{sec:qre}, they coincide exactly with 
(\ref{yuna1})--(\ref{yuna2}).
Moreover the normalization conditions (\ref{pin}) becomes (\ref{tsgmi}) 
under the correspondence (\ref{kyuki}) of $\Km$ and $\Psi$.
Therefore we conclude that the conjugation operator $\Km$ 
in our quantized reflection equation (\ref{grk}) is nothing but the 
3D $\Km$, justifying the usage of the same symbol for it.
See Appendix \ref{app:rk} for an explicit formula of $\Km$.
Here we pick some notable properties.
\begin{align}
\text{3D reflection eq.}\; & \;
\hat{\Rm}_{456}\hat{\Rm}_{489}
\Km_{3579}\hat{\Rm}_{269}\hat{\Rm}_{258}
\Km_{1678}\Km_{1234}
=\Km_{1234}
\Km_{1678}
\hat{\Rm}_{258}\hat{\Rm}_{269}
\Km_{3579}\hat{\Rm}_{489}\hat{\Rm}_{456},
\label{kyuki1}\\
\text{inversion relation}\; : \;\;
&\Km = \Km^{-1},
\label{kyuki2}\\
\text{weight conservation}\; : \;\;
& [(xy^{-1})^{{\bf h}_1}x^{{\bf h}_2}(xy)^{{\bf h}_3}y^{{\bf h}_4}, \Km]=0,
\label{kyuki3}
\end{align}
where $x, y$ are generic parameters and ${\bf h}_i$ is 
defined similarly to those in (\ref{ykn3}).
The weight conservation (\ref{kyuki3}) originates in the factor 
$\delta^{a+b+c}_{i+j+k}\,
\delta^{b+2c+d}_{j+2k+l}$ in (\ref{bd}).
The $\hat{\Rm}$ in (\ref{kyuki1}) is the 3D $\Rm$ described 
in the previous subsection.

The 3D reflection equation was proposed by Isaev and Kulish \cite{IK}.
The above solution $(\hat{\Rm},\Km)$ is due to \cite{KO1}.
In this paper we will not directly concern 
the 3D equations (\ref{ykn1}) and (\ref{kyuki1})
but rather utilize their auxiliary linear problems 
(\ref{LLLR}) and (\ref{hrk}).

\section{Reduction to Yang-Baxter equation}\label{sec:ybe}

\subsection{Concatenation of the tetrahedron equation}
Consider $n$ copies of (\ref{LLLR}) in which the 
spaces labeled with $1,2,3$ are replaced by $1_i,2_i,3_i$ with 
$i=1,2,\ldots, n$:
\begin{align*}
&(L_{1_i2_i4}L_{1_i3_i5} 
L_{2_i3_i6}) \,\Rm_{456} = 
\Rm_{456}\,(L_{2_i3_i6}L_{1_i3_i5}L_{1_i2_i4}).
\end{align*}
Sending $\Rm_{456}$ to the left by 
repeatedly applying this relation, we get
\begin{equation}\label{mikrup}
\begin{split}
&(L_{1_1 2_1 4}L_{1_1 3_1 5} 
L_{2_1 3_1 6})\cdots 
(L_{1_n 2_n 4}L_{1_n 3_n 5} 
L_{2_n 3_n 6})\,\Rm_{456} \\
&\qquad = \Rm_{456} \,(L_{2_13_16}
L_{1_13_15}
L_{1_12_14}) \cdots 
(L_{2_n 3_n 6}L_{1_n 3_n 5}
L_{1_n 2_n 4}).
\end{split}
\end{equation}
Set ${\bf V} = V^{\otimes n} \simeq (\C^2)^{\otimes n}$ in general and  
$\overset{{\bf 1}}{\bf V} 
= \overset{1_1}{V} \otimes \cdots \otimes \overset{1_n}{V}$
with labels.
The notations $\overset{{\bf 2}}{\bf V}$,  $\overset{{\bf 3}}{\bf V}$ are 
to be understood similarly.
The equality (\ref{mikrup}) holds in  
$\mathrm{End}(\overset{{\bf 1}}{\bf V} \otimes
\overset{{\bf 2}}{\bf V} \otimes
\overset{{\bf 3}}{\bf V} \otimes \overset{4}{F}_{q^2}
\otimes \overset{5}{F}_{q^2}
\otimes \overset{6}{F}_{q^2})$.

The above maneuver is just a 3D analogue 
of deriving commutation relations of monodromy 
matrices for length $n$ chain from concatenation of the local $RLL=LLR$ relation.
It is possible to rearrange (\ref{mikrup}) 
without changing the order of any two operators 
sharing common labels as
\begin{equation}\label{mikru}
\begin{split}
&(L_{1_1 2_1 4} \cdots L_{1_n 2_n 4})
(L_{1_1 3_1 5}  \cdots L_{1_n 3_n 5})
(L_{2_1 3_1 6}  \cdots L_{2_n 3_n 6})
\,\Rm_{456} \\
&\qquad = \Rm_{456} \,
(L_{2_1 3_1 6}  \cdots L_{2_n 3_n 6})
(L_{1_1 3_1 5}  \cdots L_{1_n 3_n 5})
(L_{1_1 2_1 4} \cdots L_{1_n 2_n 4}).
\end{split}
\end{equation}

\subsection{\mathversion{bold}Trace reduction}

Write (\ref{ykn3}) in the form  
$\Rm_{456}^{-1}x^{{\bf h}_4}(xy)^{{\bf h}_5} y^{{\bf h}_6} 
= x^{{\bf h}_4}(xy)^{{\bf h}_5} y^{{\bf h}_6} \Rm_{456}^{-1}$.
Multiplying this to (\ref{mikru}) from the left and
taking the trace over 
$\overset{4}{F}_{q^2}
\otimes \overset{5}{F}_{q^2}
\otimes \overset{6}{F}_{q^2}$ we get
\begin{equation}\label{mikru1}
\begin{split}
&\mathrm{Tr}_4(x^{{\bf h}_4} L_{1_1 2_1 4}\cdots
L_{1_n 2_n 4})
\mathrm{Tr}_5((xy)^{{\bf h}_5} L_{1_1 3_1 5}\cdots
L_{1_n 3_n 5})
\mathrm{Tr}_6(y^{{\bf h}_6} L_{2_1 3_1 6}\cdots
L_{2_n 3_n 6})\\
&= \mathrm{Tr}_6(y^{{\bf h}_6} L_{2_1 3_1 6}\cdots
L_{2_n 3_n 6})
\mathrm{Tr}_5((xy)^{{\bf h}_5} L_{1_1 3_1 5}\cdots
L_{1_n 3_n 5})
\mathrm{Tr}_4(x^{{\bf h}_4} L_{1_1 2_1 4}\cdots
L_{1_n 2_n 4}).
\end{split}
\end{equation}
All the factors appearing here possess the same structure as
\begin{align}\label{obata1}
S^{\mathrm {tr}}_{{\bf 1}, {\bf 2}}(z) 
= \varrho^{\mathrm {tr}}(z)
\mathrm{Tr}_a(z^{{\bf h}_a} L_{1_1 2_1 a}\cdots
L_{1_n 2_n a}) \in 
\mathrm{End}(\overset{{\bf 1}}{\bf V} \otimes \overset{{\bf 2}}{\bf V}),
\end{align}
where $a$ is a dummy label for the auxiliary Fock space $\overset{a}{F}_{q^2}$.
We have inserted a scalar 
$\varrho^{\mathrm {tr}}(z)$ which will be specified in (\ref{askS}).
Now the relation (\ref{mikru1}) is stated as the Yang-Baxter equation:
\begin{equation}\label{ybe1}
S^{\mathrm {tr}}_{{\bf 1}, {\bf 2}}(x)
S^{\mathrm {tr}}_{{\bf 1}, {\bf 3}}(xy)
S^{\mathrm {tr}}_{{\bf 2}, {\bf 3}}(y)
=
S^{\mathrm {tr}}_{{\bf 2}, {\bf 3}}(y)
S^{\mathrm {tr}}_{{\bf 1}, {\bf 3}}(xy)
S^{\mathrm {tr}}_{{\bf 1}, {\bf 2}}(x).
\end{equation}
In the sequel, we will 
often suppress the labels of the spaces like ${\bf 1}, {\bf 2}$ etc 
without notice if they are unnecessary.
The above construction of $S^{\mathrm {tr}}(z)$ is 
due to \cite{BS}, where
it was claimed to yield the quantum $R$ matrix 
for the antisymmetric tensor representations of $U_p(A^{(1)}_{n-1})$
with some $p$. 
A precise description adapted to the present convention 
is available in Appendix \ref{app:str}.

\subsection{\mathversion{bold}Boundary vector reduction}\label{ss:bbr}

For $s=1,2$, introduce the vectors 
\begin{align}
&\langle \chi_s| = \sum_{m\ge 0}\frac{\langle sm|}{(q^{2s^2})_m} 
\in F^\ast_{q^2},\qquad\,
|\chi_s\rangle = \sum_{m\ge 0}\frac{|sm\rangle}{(q^{2s^2})_m}
\in F_{q^2},
\label{xk}\\
&\langle\eta_s| = \sum_{m\ge 0}\frac{\langle sm|}{(q^{s^2})_m} \in F^\ast_q,
\qquad\quad
|\eta_s\rangle = \sum_{m\ge 0}\frac{|sm\rangle}{(q^{s^2})_m} \in F_q.
\label{xb}
\end{align}
They are characterized 
by $|\chi_s\rangle = |\eta_s\rangle|_{q\rightarrow q^2}$, 
$\langle \chi_s| = \langle \eta_s||_{q\rightarrow q^2}$ and a
kind of coherent-vector like condition:
\begin{equation*}
\begin{split}
\apm |\eta_1\rangle &= (1 \mp q^{\mp \hf}\ok)|\eta_1\rangle,
\qquad\,
\langle \eta_1| \apm = \langle \eta_1| (1\pm q^{\pm \hf}\ok),\\
\ap | \eta_2\rangle &= \am  | \eta_2\rangle, \qquad\qquad\qquad
\langle \eta_2 | \ap = \langle \eta_2 | \am.
\end{split}
\end{equation*}
The following relations are proved in \cite[Prop.4.1]{KS}: 
\begin{align}\label{serge}
(\langle \chi_s| \otimes 
\langle \chi_s| \otimes \langle \chi_s| )\Rm = 
\langle \chi_s| \otimes 
\langle \chi_s| \otimes \langle \chi_s| ,\quad
\Rm (|\chi_s\rangle \otimes|\chi_s\rangle \otimes|\chi_s\rangle)
=  |\chi_s\rangle \otimes|\chi_s\rangle \otimes|\chi_s\rangle.
\end{align}
Sandwich the relation  (\ref{mikru}) 
between the bra vector 
$(\langle \overset{4}{\chi_s}| \otimes
\langle \overset{5}{\chi_s}| \otimes
\langle \overset{6}{\chi_s}|)
x^{{\bf h}_4} (xy)^{{\bf h}_5}y^{{\bf h}_6}$ and the 
ket vector 
$|\overset{4}{\chi_{s'}}\rangle 
\otimes |\overset{5}{\chi_{s'}}\rangle  \otimes 
|\overset{6}{\chi_{s'}}\rangle$.
Using (\ref{serge}) and (\ref{ykn3}) we find
\begin{equation}\label{mikru2}
\begin{split}
&\langle\overset{4}{\chi_s}| x^{{\bf h}_4}
L_{1_1 2_1 4}\cdots
L_{1_n 2_n 4}|\overset{4}{\chi_{s'}}\rangle
\langle\overset{5}{\chi_s}| (xy)^{{\bf h}_5}
L_{1_1 3_1 5}\cdots
L_{1_n 3_n 5}|\overset{5}{\chi_{s'}}\rangle
\langle\overset{6}{\chi_s}| y^{{\bf h}_6}
L_{2_1 3_1 6}\cdots
L_{2_n 3_n 6}|\overset{6}{\chi_{s'}}\rangle\\
&=
\langle\overset{6}{\chi_s}| y^{{\bf h}_6}
L_{2_1 3_1 6}\cdots
L_{2_n 3_n 6}|\overset{6}{\chi_{s'}}\rangle
\langle\overset{5}{\chi_s}| (xy)^{{\bf h}_5}
L_{1_1 3_1 5}\cdots
L_{1_n 3_n 5}|\overset{5}{\chi_{s'}}\rangle
\langle\overset{4}{\chi_s}| x^{{\bf h}_4}
L_{1_1 2_1 4}\cdots
L_{1_n 2_n 4}|\overset{4}{\chi_{s'}}\rangle.
\end{split}
\end{equation}
All the factors appearing here have the form
\begin{align}\label{obata2}
S^{s, s'}_{{\bf 1}, {\bf 2}}(z) 
= \varrho^{s,s'}(z)
\langle \overset{a}{\chi_s}| 
z^{{\bf h}_a} L_{1_1 2_1 a}\cdots
L_{1_n 2_n a}
|\overset{a}{\chi_{s'}} \rangle\in 
\mathrm{End}(\overset{{\bf 1}}{\bf V} \otimes \overset{{\bf 2}}{\bf V})
\qquad  (s, s'=1,2),
\end{align}
where the notation is similar to (\ref{obata1}).
The scalar $\varrho^{s, s'}(z)$ will be specified in (\ref{askS}).
Now (\ref{mikru2}) is stated as the Yang-Baxter equation:
\begin{equation}\label{ybe2}
S^{s, s'}_{{\bf 1}, {\bf 2}}(x)
S^{s, s'}_{{\bf 1}, {\bf 3}}(xy)
S^{s, s'}_{{\bf 2}, {\bf 3}}(y)
=
S^{s, s'}_{{\bf 2}, {\bf 3}}(y)
S^{s, s'}_{{\bf 1}, {\bf 3}}(xy)
S^{s, s'}_{{\bf 1}, {\bf 2}}(x).
\end{equation}
This construction of the four solutions 
corresponding to the choice 
$1\le s, s' \le 2$
is due to \cite{KS}, where the cases 
$(s, s')=(1,1), (2,1)$ and $(2,2)$
were identified with the quantum $R$ matrices  
for the spin representations of 
$U_p(D^{(2)}_{n+1}), U_p(B^{(1)}_n)$ and $U_p(D^{(1)}_n)$ with some $p$.
See Appendix \ref{app:str} for a precise description 
including the case $(s,s')=(1,2)$.

\subsection{\mathversion{bold}Basic properties of 
$S^{\mathrm {tr}}(z)$ and $S^{s,s'}(z)$}\label{ss:bss}

We write the base vectors of ${\bf V} = V^{\otimes n}$ as
$|\alb\rangle= v_{\alpha_1} \otimes \cdots \otimes v_{\alpha_n}$
in terms of an array 
$\alb=(\alpha_1,\ldots,\alpha_n) \in \{0,1\}^n$\footnote{
$|\alb\rangle \in {\bf V}$ should not be confused with 
the base $|m\rangle$ of the Fock space
containing a single integer.}. 
Set
\begin{align*}
&S(z)(|\alb\rangle \otimes |\beb\rangle)
= \sum_{\gab,\deb \in \{0,1\}^n}
S(z)_{\alb,\beb}^{\gab,\deb}
|\gab\rangle \otimes |\deb\rangle\qquad 
(S= S^{\mathrm {tr}}, S^{s,s'}).
\end{align*}
Then the formulas (\ref{obata1}) and (\ref{obata2}) 
imply the matrix product structure as
\begin{align}
&S^{\mathrm {tr}}(z)_{\alb,\beb}^{\gab,\deb}
= \varrho^{\mathrm {tr}}(z)
\mathrm{Tr}\bigl(z^{{\bf h}}
L^{\gamma_1,\delta_1}_{\alpha_1, \beta_1}
\cdots
L^{\gamma_n,\delta_n}_{\alpha_n, \beta_n}\bigr),
\label{strz}
\\
&S^{s,s'}(z)_{\alb,\beb}^{\gab,\deb}
= \varrho^{s,s'}(z)
\langle \chi_s |z^{{\bf h}}
L^{\gamma_1,\delta_1}_{\alpha_1, \beta_1}
\cdots
L^{\gamma_n,\delta_n}_{\alpha_n, \beta_n}|\chi_{s'}\rangle,
\label{s11}
\end{align}
where $L^{\gamma,\delta}_{\alpha,\beta}$ is 
given by (\ref{air}) and $\mathrm{Tr}(\cdots)$ and 
$\langle \chi_s| (\cdots) |\chi_{s'}\rangle$ are 
taken over $F_{q^2}$.
They are evaluated by using the commutation relations (\ref{ngh2}), 
the formulas in (\ref{lin}) with 
$q \rightarrow q^2$ and 
\begin{align}
&\mathrm{Tr}(z^{\bf h} \OK^r(\Ap)^s(\Am)^{s'}) = 
\delta_{s,s'}\frac{q^r(q^4;q^4)_s}{(zq^{2r};q^4)_{s+1}}.
\label{yuk1}
\end{align}

The expression (\ref{s11}) is shown diagrammatically
as the barbeque stick with $n$ X-shape sausage\footnote{
Each sausage carries $V\otimes V$ 
whereas the stick does $F_{q^2}$.  
The trace (\ref{strz}) corresponds to a ring shape stick. 
Description due to Sergey Sergeev.} 
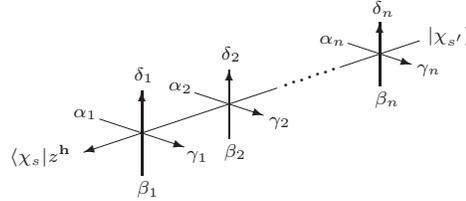
\[
\begin{picture}(200,75)(-100,-40)
\put(3,1){\vector(-3,-1){73}}

\put(-98,-27){$\scriptstyle{\langle \chi_s|z^{\bf h}}$}

\put(-48,-16){\vector(0,1){16}}\put(-48,-16){\line(0,-1){16}}
\put(-48,-16){\vector(3,-1){16}} \put(-48,-16){\line(-3,1){16}}
\put(-51,4){$\scriptstyle{\delta_1}$}
\put(-74,-10){$\scriptstyle{\alpha_1}$}
\put(-31,-26){$\scriptstyle{\gamma_1}$}
\put(-50,-39){$\scriptstyle{\beta_1}$}

\put(-15,-5){\vector(0,1){13}}\put(-15,-5){\line(0,-1){13}}
\put(-15,-5){\vector(3,-1){13}}\put(-15,-5){\line(-3,1){13}}
\put(-38,0){$\scriptstyle{\alpha_2}$}
\put(-18,12){$\scriptstyle{\delta_2}$}
\put(-17,-25){$\scriptstyle{\beta_2}$}
\put(0,-13){$\scriptstyle{\gamma_2}$}

\multiput(5.1,1.7)(3,1){7}{.} 
\put(6,2){
\put(21,7){\line(3,1){30}}
\put(36,12){\vector(0,1){12}}\put(36,12){\line(0,-1){12}}
\put(36,12){\vector(3,-1){12}}\put(36,12){\line(-3,1){12}}
\put(13,16){$\scriptstyle{\alpha_n}$}
\put(50,5){$\scriptstyle{\gamma_n}$}
\put(33,27){$\scriptstyle{\delta_n}$}
\put(34,-7){$\scriptstyle{\beta_n}$}

}
 
\put(60,19){$\scriptstyle{|\chi_{s'} \rangle}$}
\end{picture}
\]
In view of this, we call (\ref{xk}) and (\ref{xb}) 
the {\em boundary vectors}, 
and the procedure in Section \ref{ss:bbr} the {\em boundary vector reduction}.
Intriguingly the boundary vectors are known to reflect the 
end shape of the Dynkin diagrams of the relevant 
quantum affine algebras. See Appendix \ref{app:str} and \cite[Rem.7.2]{KS}. 
The vectors $\langle\eta_s|$ and $|\eta_s\rangle$ will play a similar role  
for the quantized reflection equation in Section \ref{sec:re}.

For $\alb=(\alpha_1,\ldots, \alpha_n) \in \{0,1\}^n$ set 
\begin{align}\label{askB}
|\alb| =\alpha_1+\cdots + \alpha_n,\qquad
{\bf V}_k = \bigoplus_{\alb \in \{0,1\}^n, \, |\alb | =k}\!\!\!
\C |\alb\rangle, \qquad
{\bf V}^\pm = \bigoplus_{\alb \in \{0,1\}^n, \,  (-1)^{|\alb |} = \pm 1}\!\!\!
\C |\alb\rangle.
\end{align}
By the definition the following direct sum decomposition holds:
\begin{align*}
{\bf V}= V^{\otimes n} 
&= {\bf V}_0 \oplus {\bf V}_1 \oplus \cdots \oplus {\bf V}_n,
\qquad
{\bf V} = {\bf V}^+ \oplus {\bf V}^-.
\end{align*}
From (\ref{misaki40}), (\ref{mzsma1}), (\ref{mzsma2}) 
and (\ref{xk}) one can show
\begin{align}
&S^{\mathrm{tr}}(z)_{\alb,\beb}^{\gab,\deb}
=z^{|\beb|-|\deb|}
S^{\mathrm{tr}}(z)^{\alb^\vee,\beb^\vee}_{\gab^\vee,\deb^\vee},
\quad
S^{s,s'}(z)_{\alb,\beb}^{\gab,\deb}
=z^{|\beb|-|\deb|}
S^{s', s}(z)^{\alb^\vee,\beb^\vee}_{\gab^\vee,\deb^\vee},
\label{yume15}\\
&S(z)_{\alb,\beb}^{\gab,\deb}=0 \;\; \text{unless}\;\;
\alb + \beb = \gab + \deb \in \Z^n
\qquad (S= S^{\mathrm {tr}}, S^{s,s'}),
\label{yume2}\\
&S^{\mathrm {tr}}(z)_{\alb,\beb}^{\gab,\deb}=0 
\;\; \text{unless}
\;\; |\alb| = |\gab| \;\; 
\text{and}\;\; |\beb| = |\deb|,
\label{yume3}\\
&S^{2,2}(z)_{\alb,\beb}^{\gab,\deb}=0 
\;\; \text{unless}
\;\; |\alb| \equiv  |\gab| \;\; 
\text{and}\;\; |\beb| \equiv |\deb|\;\mod 2,
\label{yume4}
\end{align}
where $\alb^\vee = (\alpha_n,\ldots, \alpha_1)$ signifies the 
reversal of the array $\alb=(\alpha_1,\ldots, \alpha_n)$.
The properties (\ref{yume3}) and (\ref{yume4})
imply the decomposition
\begin{align}
S^{\mathrm {tr}}(z) &= \bigoplus_{0 \le l,m \le n}
S^{\mathrm {tr}}_{l,m}(z),\qquad\;\,
S^{\mathrm {tr}}_{l,m}(z) \in \mathrm{End}({\bf V}_l\otimes {\bf V}_m),
\label{nami}\\
S^{2,2}(z) &= \bigoplus_{\sigma, \sigma'= +, -}
S^{2,2}_{\sigma, \sigma'}(z),
\qquad
S^{2,2}_{\sigma, \sigma'}(z) 
\in \mathrm{End}({\bf V}^\sigma \otimes {\bf V}^{\sigma'}).
\label{nami2}
\end{align}
 
The Yang-Baxter equations (\ref{ybe1}) and $(\ref{ybe2})|_{s=s'=2}$ are valid 
within the subspaces
${\bf V}_k \otimes {\bf V}_l \otimes {\bf V}_m$ and  
${\bf V}^{\sigma} \otimes {\bf V}^{\sigma'} \otimes {\bf V}^{\sigma''}$
of 
$\overset{{\bf 1}}{\bf V} \otimes 
\overset{{\bf 2}}{\bf V} \otimes
\overset{{\bf 3}}{\bf V}$, respectively.
The scalar prefactors in (\ref{strz}) and 
(\ref{s11}) may be specified 
depending on the summands in (\ref{nami}) and (\ref{nami2}). 
We take them as
\begin{equation}\label{askS}
\begin{split}
&\varrho^{\mathrm {tr}}_{l,m}(z) =
q^{-|l-m|}(1-z q^{2|l-m|}),
\qquad
\varrho^{s,s'}(z) =
\frac{(z^u;q^{2ss'})_\infty}
{(-z^uq^2;q^{2ss'})_\infty}\quad ((s,s') \neq (2,2)),\\
&\varrho^{2,2}_{\pm,\pm}(z)= 
\frac{(z^2;q^8)_\infty}{(z^2q^4;q^8)_\infty},\qquad
\varrho^{2,2}_{\pm,\mp}(z)= 
\frac{(z^2q^4;q^8)_\infty}{q(z^2q^8;q^8)_\infty},
\end{split}
\end{equation}
where $u=\max(s,s')$.
This choice makes all the matrix elements of 
$S^{\mathrm {tr}}_{l,m}(z)$ and $S^{s,s'}(z)$ 
rational functions of $z$ and $q$.
It also simplifies some ``typical" elements so that
\begin{equation}\label{askS1}
\begin{split}
&S^{\mathrm {tr}}_{l,m}(z) (|{\bf e}_1+\cdots + {\bf e}_l\rangle
\otimes |{\bf e}_1+\cdots + {\bf e}_m\rangle)
=(-1)^{\max(l-m,0)} |{\bf e}_1+\cdots + {\bf e}_l\rangle 
\otimes |{\bf e}_1+\cdots + {\bf e}_m\rangle,
\\
&S(z) (|{\bf 0} \rangle \otimes |{\bf 0} \rangle)
= |{\bf 0} \rangle \otimes |{\bf 0} \rangle \quad 
(S=S^{1,1}, S^{1,2},S^{2,1}, S^{2,2}_{+,+}),
\qquad
S^{2,2}_{-,-}(z) (|{\bf e}_1 \rangle \otimes |{\bf e}_1 \rangle)
=|{\bf e}_1 \rangle \otimes |{\bf e}_1 \rangle,
\\
&S^{2,2}_{+,-}(z)(|{\bf 0}\rangle \otimes |{\bf e}_1\rangle) 
= |{\bf 0}\rangle \otimes |{\bf e}_1\rangle,
\qquad
S^{2,2}_{-,+}(z)(|{\bf e}_1\rangle \otimes |{\bf 0}\rangle) 
= -|{\bf e}_1\rangle \otimes |{\bf 0}\rangle.
\end{split}
\end{equation}

\section{Reduction to reflection equation}\label{sec:re}

Starting from the quantized reflection equation (\ref{grk}),
one can perform two kinds of reductions similar to Section \ref{sec:ybe} to 
construct the 2D $K$ matrices systematically in the matrix product form.
This is the main result of the paper which we are going to present in this section.

\subsection{Concatenation of quantized reflection equation}

Consider $n$ copies of (\ref{hrk}) in which the spaces labeled with $1,2$ are
replaced by $1_i, 2_i$ with $i=1,2,\ldots, n$:
\begin{align}\label{kanon}
L_{1_i 2_i 3}K_{2_i 4}
L_{2_i 1_i 5}K_{1_i 6}\Km_{3456} = 
\Km_{3456}K_{1_i 6}L_{1_i 2_i 5}
K_{2_i 4}L_{2_i 1_i 3}.
\end{align}
As commented after (\ref{yuna2}) we know that the 3D $\Km$ 
characterized by (\ref{reK}) and detailed in (\ref{bd}), (\ref{rest})
makes this relation hold.
Using (\ref{kanon}) successively, one can let $\Km_{3456}$ penetrate 
$L_{1_i 2_i 3}K_{2_i 4}
L_{2_i 1_i 5}K_{1_i 6}$ through to the left 
converting it into $K_{1_i 6}L_{1_i 2_i 5}
K_{2_i 4}L_{2_i 1_i 3}$
$(i=1,2,\ldots, n)$ as
\begin{equation*}
\begin{split}
&(L_{1_1 2_1 3}K_{2_1 4}
L_{2_1 1_1 5}K_{1_1 6})\cdots
(L_{1_n 2_n 3}K_{2_n 4}
L_{2_n 1_n 5}K_{1_n 6})\Km_{3456}\\
&\qquad = \Km_{3456}(K_{1_1 6}L_{1_1 2_1 5}
K_{2_1 4}L_{2_1 1_1 3})
\cdots
(K_{1_n 6}L_{1_n 2_n 5}
K_{2_n 4}L_{2_n 1_n 3}).
\end{split}
\end{equation*}
One can rearrange this without changing the order of operators 
sharing common labels as
\begin{equation}\label{kanon1}
\begin{split}
&(L_{1_1 2_1 3} \cdots L_{1_n 2_n 3})
(K_{2_1 4} \cdots K_{2_n 4})
(L_{2_1 1_1 5} \cdots L_{2_n 1_n 5})
(K_{1_1 6} \cdots K_{1_n 6})
\Km_{3456}\\
&\qquad = \Km_{3456}
(K_{1_1 6} \cdots K_{1_n 6})
(L_{1_1 2_1 5} \cdots L_{1_n 2_n 5})
(K_{2_1 4} \cdots K_{2_n 4})
(L_{2_1 1_1 3} \cdots L_{2_n 1_n 3}).
\end{split}
\end{equation}

\subsection{Trace reduction}

Write (\ref{kyuki3}) as 
$\Km^{-1}_{3456}
(xy^{-1})^{{\bf h}_3}x^{{\bf h}_4}(xy)^{{\bf h}_5}y^{{\bf h}_6}
= 
(xy^{-1})^{{\bf h}_3}x^{{\bf h}_4}(xy)^{{\bf h}_5}y^{{\bf h}_6}\Km^{-1}_{3456}$.
Multiplying this to (\ref{kanon1}) from the left and taking the trace over 
$\overset{3}{F}_{q^2}\otimes
\overset{4}{F}_{q} \otimes
\overset{5}{F}_{q^2} \otimes
\overset{6}{F}_{q}$,
we obtain
\begin{equation}\label{kanon2}
\begin{split}
&\mathrm{Tr}_3\bigl((xy^{-1})^{{\bf h}_3} L_{1_1 2_1 3}
\cdots L_{1_n 2_n 3}\bigr)
\mathrm{Tr}_4\bigl(x^{{\bf h}_4}K_{2_1 4} 
\cdots K_{2_n 4}\bigr) \times \\
&\qquad\qquad
\times \mathrm{Tr}_5\bigl((xy)^{{\bf h}_5} L_{2_1 1_1 5}
\cdots L_{2_n 1_n 5}\bigr)
\mathrm{Tr}_6\bigl(y^{{\bf h}_6}K_{1_1 6} 
\cdots K_{1_n 6}\bigr)\\
&= \mathrm{Tr}_6\bigl(y^{{\bf h}_6}
K_{1_1 6} \cdots K_{1_n 6}\bigr)
\mathrm{Tr}_5\bigl((xy)^{{\bf h}_5}
L_{1_1 2_1 5} \cdots 
L_{1_n 2_n 5}\bigr) \times \\
&\qquad \qquad \times
\mathrm{Tr}_4\bigl(x^{{\bf h}_4}
K_{2_1 4} \cdots K_{2_n 4}\bigr)
\mathrm{Tr}_3\bigl((xy^{-1})^{{\bf h}_3}
L_{2_1 1_1 3} \cdots L_{2_n 1_n 3}\bigr).
\end{split}
\end{equation}
Here $\mathrm{Tr}_3(\cdots)$ and $\mathrm{Tr}_5(\cdots)$  
are identified with $S^{\mathrm{tr}}(z)$ in (\ref{obata1}). 
The other factors emerging from $K$ have the form
\begin{align}\label{sae}
K^{\mathrm{tr}}_{\bf 1}(z) 
= \kappa^{\mathrm{tr}}(z)
\mathrm{Tr}_a\bigl(z^{{\bf h}_a}
K_{1_1 a} \cdots K_{1_n a}\bigr)
\in \mathrm{End}(\overset{\bf 1}{\bf V}),
\end{align}
where 
$\overset{{\bf 1}}{\bf V} = 
\overset{1_1}{V} \otimes \cdots \otimes \overset{1_n}{V}
\simeq (\C^2)^{\otimes n}$ 
as before.
The trace is taken over $\overset{a}{F}_q$ and evaluated 
by means of (\ref{ngh1}) and $(\ref{yuk1})|_{q\rightarrow q^{1/2}}$.
The scalar $\kappa^{\mathrm{tr}}(z)$ will be specified in (\ref{sren1}).
Now the relation (\ref{kanon2}) 
is the usual reflection equation in 2D:
\begin{align}\label{misakiB}
S^{\mathrm{tr}}_{{\bf 1}, {\bf 2}}(xy^{-1})
K^{\mathrm{tr}}_{\bf 2}(x) 
S^{\mathrm{tr}}_{{\bf 2},{\bf 1}}(xy)
K^{\mathrm{tr}}_{\bf 1}(y) 
=
K^{\mathrm{tr}}_{\bf 1}(y) 
S^{\mathrm{tr}}_{{\bf 1},{\bf 2}}(xy)
K^{\mathrm{tr}}_{\bf 2}(x) 
S^{\mathrm{tr}}_{{\bf 2}, {\bf 1}}(xy^{-1}).
\end{align}

The construction (\ref{sae}) implies the matrix product formula for each element as
\begin{equation}\label{misato}
\begin{split}
K^{\mathrm{tr}}(z) |\alb\rangle &= \sum_{\beb \in \{0,1\}^n}
K^{\mathrm{tr}}(z)_{\alb}^{\beb}|\beb\rangle,\\
K^{\mathrm{tr}}(z)_{\alb}^{\beb} 
& = \kappa^{\mathrm{tr}}(z)
\mathrm{Tr}\bigl(z^{{\bf h}}
K^{\beta_1}_{\alpha_1}
\cdots 
K^{\beta_n}_{\alpha_n}
\bigr)
\end{split}
\end{equation}
in terms of $K^{\beta}_{\alpha}$ 
specified in (\ref{air2}).

To derive the selection rule of (\ref{misato}), 
suppose the number of pairs 
$(0,0)$, 
$(0,1)$, 
$(1,0)$, 
$(1,1)$ in the multiset 
$\{(\alpha_1,\beta_1), \ldots, 
(\alpha_n, \beta_n)\}$ is
$r, s, t, u$,
respectively in (\ref{misato}).
By the definition (\ref{askB}) we have 
$|\alb| = t+u$, 
$|\beb| = s+u$ 
and 
$n = r + s + t + u$.
Moreover in order to have a non-vanishing matrix element (\ref{misato}),
there must be as many creation operators as annihilation operators.
From (\ref{air2}) or (\ref{kfig}), 
this imposes the constraint $r=u$.
These relations force 
$|\alb| + |\beb| = n$.
Namely we have the ``dual weight" conservation: 
\begin{align}\label{misato2}
K^{\mathrm{tr}}(z)_{\alb}^{\beb} = 0\quad\text{unless}\;\;
|\alb| + |\beb| = n
\end{align}
or equivalently, the direct sum decomposition:
\begin{align}\label{misato3}
K^{\mathrm{tr}}(z) = \bigoplus_{0 \le l \le n} 
K^{\mathrm{tr}}_l(z),\qquad
K^{\mathrm{tr}}_l(z): \,{\bf V}_l \rightarrow {\bf V}_{n-l}.
\end{align}
The space ${\bf V}_l$ (\ref{askB}) is naturally regarded as 
a fundamental $U_p(A^{(1)}_{n-1})$-module. See Appendix \ref{app:str}.
The above result suggests that
a natural representation theoretical formulation 
of the boundary reflection is ${\bf V}_l \rightarrow {\bf V}_{n-l}$  
rather than ${\bf V}_l \rightarrow {\bf V}_l$.

The equality (\ref{misakiB}) holds in a finer manner, i.e.
as the identity of linear operators 
${\bf V}_l \otimes {\bf V}_m 
\rightarrow {\bf V}_{n-l}\otimes {\bf V}_{n-m}$ for each 
pair $(l,m) \in \{0,1,\ldots, n\}^2$.
The scalar in (\ref{sae})  can be specified 
depending on $l$ as $\kappa^{\mathrm{tr}}_l(z)$.
We take it as
\begin{align}\label{sren1}
\kappa^{\mathrm{tr}}_l(z) = (-1)^lq^{-\frac{n}{2}}
(1-z q^n).
\end{align}
Then applying (\ref{kfig}) and 
$(\ref{yuk1})|_{q\rightarrow q^{1/2}}$ 
to (\ref{misato}),  it is easy to check
\begin{align*}
K^{\mathrm{tr}}_l(z)|{\bf e}_1+\cdots + {\bf e}_l\rangle = 
|{\bf e}_{l+1}+\cdots + {\bf e}_n\rangle + \cdots
\qquad (0 \le l \le n).
\end{align*}

\subsection{Boundary vector reduction}

Supported by computer experiments we conjecture 
\begin{equation}\label{syki}
\begin{split}
(\langle \chi_s| \otimes 
\langle \eta_k | \otimes 
\langle \chi_s | \otimes
\langle \eta_k| ) \Km &= \langle \chi_s| \otimes 
\langle \eta_k | \otimes 
\langle \chi_s | \otimes
\langle \eta_k|\qquad (1 \le s \le k \le 2),
\\
\Km (|\chi_s\rangle \otimes
|\eta_k \rangle \otimes
|\chi_s \rangle \otimes
|\eta_k\rangle)
&= |\chi_s\rangle \otimes
|\eta_k \rangle \otimes
|\chi_s \rangle \otimes
|\eta_k\rangle\qquad (1 \le s \le k \le 2),
\end{split}
\end{equation}
where 
the components are defined in (\ref{xk}) and (\ref{xb}).
Sandwich the relation (\ref{kanon1}) 
between the bra vector
$(
\langle \overset{3}{\chi_s}| \otimes 
\langle \overset{4}{\eta_k}| \otimes 
\langle \overset{5}{\chi_s}| \otimes
\langle \overset{6}{\eta_k}|)
(xy^{-1})^{{\bf h}_3}x^{{\bf h}_4}(xy)^{{\bf h}_5}y^{{\bf h}_6}$ and 
the ket vector 
$|\overset{3}{\chi_{s'}}\rangle \otimes
|\overset{4}{\eta_{k'}}\rangle \otimes
|\overset{5}{\chi_{s'}}\rangle \otimes
|\overset{6}{\eta_{k'}} \rangle$.
Thanks to (\ref{syki}) the result reduces to
\begin{equation}\label{sizka0}
\begin{split}
&\langle \overset{3}{\chi_s}|(xy^{-1})^{{\bf h}_3}
L_{1_1 2_1 3} \cdots 
L_{1_n 2_n 3}|\overset{3}{\chi_{s'}}\rangle
\langle \overset{4}{\eta_k}|x^{{\bf h}_4}
K_{2_1 4}\cdots K_{2_n 4}
|\overset{4}{\eta_{k'}}\rangle \times\\
&\qquad\quad
\times \langle \overset{5}{\chi_s}|(xy)^{{\bf h}_5}
L_{2_1 1_1 5}\cdots 
L_{2_n 1_n 5} |\overset{5}{\chi_{s'}}\rangle
\langle \overset{6}{\eta_k}|y^{{\bf h}_6}
K_{1_1 6}\cdots K_{1_n 6}
|\overset{6}{\eta_{k'}}\rangle\\
&= 
\langle \overset{6}{\eta_k}|y^{{\bf h}_6}
K_{1_1 6}\cdots K_{1_n 6}
|\overset{6}{\eta_{k'}}\rangle
\langle \overset{5}{\chi_{s}}|(xy)^{{\bf h}_5}
L_{1_1 2_1 5}\cdots 
L_{1_n 2_n 5} |\overset{5}{\chi_{s'}}\rangle \times \\
&\qquad\quad
\times \langle \overset{4}{\eta_k}|x^{{\bf h}_4}
K_{2_1 4}\cdots K_{2_n 4}
|\overset{4}{\eta_{k'}}\rangle
\langle \overset{3}{\chi_s}|(xy^{-1})^{{\bf h}_3}
L_{2_1 1_1 3} \cdots 
L_{2_n 1_n 3}|\overset{3}{\chi_{s'}}\rangle.
\end{split}
\end{equation}
The factors 
$\langle \chi_s|(\cdots)| \chi_{s'}\rangle$ involving $L$ 
are identified with $S^{s,s'}(z)$ in (\ref{obata2}).
The other factors emerging from $K$ have the form
\begin{align}\label{sizka}
K^{k,k'}_{\bf 1}(z) = \kappa^{k,k'}(z)
\langle \overset{a}{\eta_k}|
z^{{\bf h}_a}
K_{1_1 a}\cdots K_{1_n a}
|\overset{a}{\eta_{k'}}\rangle \in \mathrm{End}(\overset{\bf 1}{\bf V})
\qquad  (k, k'=1,2),
\end{align}
where the scalar $\kappa^{k,k'}(z)$ will be specified in (\ref{marine}).
The quantities $\langle \eta_k|(\cdots)|\eta_{k'}\rangle$ are evaluated 
by means of (\ref{ngh1}) and the following formulas:
\begin{equation}\label{lin}
\begin{split}
&\langle \eta_k|z^{\bf h} (\apm)^j \ok^m 
w^{\bf h}|\eta_{k'}\rangle =
\langle \eta_{k'}|w^{\bf h}\ok^m  (\amp)^j 
z^{\bf h}|\eta_k\rangle\quad (k, k' = 1,2),
\\
&\langle \eta_1|z^{\bf h}(\ap)^j \ok^m 
w^{\bf h}|\eta_1\rangle
= q^{\frac{m}{2}}z^j(-q;q)_j
\frac{(-q^{j+m+1}zw;q)_\infty}{(q^mzw;q)_\infty},
\\
&\langle \eta_1|z^{\bf h}(\am)^j \ok^m 
w^{\bf h}|\eta_2\rangle
= q^{\frac{m}{2}}z^{-j}\sum_{i=0}^j(-1)^i q^{\frac{1}{2}i(i+1-2j)}
\binom{j}{i}_{\!\!q}
\frac{(-q^{2i+2m+1}z^2w^2;q^2)_\infty}{(q^{2i+2m}z^2w^2;q^2)_\infty},
\\
&\langle \eta_1|z^{\bf h} (\ap)^j \ok^m 
w^{\bf h}|\eta_2\rangle
= q^{\frac{m}{2}}z^{j}\sum_{i=0}^j q^{\frac{1}{2}i(i+1)}
\binom{j}{i}_{\!\!q}
\frac{(-q^{2i+2m+1}z^2w^2;q^2)_\infty}{(q^{2i+2m}z^2w^2;q^2)_\infty},
\\
&\langle \eta_2|z^{\bf h} (\ap)^{j} \ok^m 
w^{\bf h}|\eta_2\rangle
= \theta(j\in 2\Z) \,q^{\frac{m}{2}}z^{j}(q^2;q^4)_{j/2}
\frac{(q^{2j+2m+2}z^2w^2;q^4)_\infty}{(q^{2m}z^2w^2;q^4)_\infty}.
\end{split}
\end{equation}
These are easily derived by only using the elementary identity
\begin{align*}
\sum_{j\ge 0}\frac{(w;q)_j}{(q;q)_j}z^j 
= \frac{(w z;q)_\infty}{(z;q)_\infty}.
\end{align*}

In terms of (\ref{sizka}) and (\ref{s11}), 
the relation (\ref{sizka0}) is stated as the reflection equation:
\begin{align}\label{aoi}
S^{s,s'}_{{\bf 1}, {\bf 2}}(xy^{-1})
K^{k,k'}_{\bf 2}(x)
S^{s,s'}_{{\bf 2},{\bf 1}}(xy)
K^{k,k'}_{\bf 1}(y)
=
K^{k,k'}_{\bf 1}(y)
S^{s,s'}_{{\bf 1},{\bf 2}}(xy)
K^{k,k'}_{\bf 2}(x)
S^{s,s'}_{{\bf 2}, {\bf 1}}(xy^{-1})
\end{align}
for any $1 \le s \le k \le 2$ and $1 \le s' \le k' \le 2$.
Thus we get, assuming (\ref{syki}),  the solutions 
$(S^{s,s'}(z), K^{k,k'}(z))$ to the reflection equation
involving the quantum $R$ matrices for the spin representation of 
$U_p(D^{(2)}_{n+1})$, $U_p(B^{(1)}_{n})$, $U_p(\tilde{B}^{(1)}_{n})$ and 
$U_p(D^{(1)}_{n})$.

The construction (\ref{sizka}) implies the matrix product formula for each element as 
\begin{equation}\label{minami}
\begin{split}
K^{k,k'}(z) |\alb\rangle &= \sum_{\beb \in \{0,1\}^n}
K^{k,k'}(z)_{\alb}^{\beb}|\beb\rangle,\\
K^{k,k'}(z)_{\alb}^{\beb} 
&= \kappa^{k,k'}(z)
\langle \eta_k| z^{{\bf h}}
K^{\beta_1}_{\alpha_1}
\cdots 
K^{\beta_n}_{ \alpha_n}
|\eta_{k'}\rangle
\end{split}
\end{equation}
in terms of $K^{\beta}_{\alpha}$ in (\ref{air2}).
From (\ref{misaki40}) and the fact 
$\kappa^{k,k'}(z) = \kappa^{k',k}(z)$ in (\ref{marine}), it can be shown that
\begin{align}\label{aimi2}
K^{k,k'}(z)^{\beb}_{\alb} = 
z^{n-|\alb|-|\beb|}
K^{k',k}(z)^{{\bf e}_1+\cdots 
+ {\bf e}_n-\alb^\vee}_{{\bf e}_1+\cdots + {\bf e}_n-\beb^\vee},
\end{align}
where $\vee$ is the same as in (\ref{yume15}). 
Noting the factor $\theta(j\in 2\Z)$ in the last formula in (\ref{lin}),
one can show 
\begin{align}\label{minami2}
K^{2,2}(z)_{\alb}^{\beb} = 0\quad\text{unless}\;\;
|\alb| + |\beb| \equiv  n \mod 2
\end{align}
by an argument similar to that given after (\ref{misato}). 
Consequently the direct sum decomposition
\begin{align*}
K^{2,2}(z) &= K^{2,2}_+(z) \oplus K^{2,2}_-(z),\qquad\quad
K^{2,2}_\sigma(z): {\bf V}^\sigma \rightarrow {\bf V}^{\sigma(-1)^n}
\end{align*}
holds, where ${\bf V}^\pm$ was defined in (\ref{askB}).
As for $K^{k,k'}(z)$ with $(k,k')\neq (2,2)$,
there is no selection rule like (\ref{misato2}) nor (\ref{minami2}).
We choose the scalar $\kappa^{k,k'}(z)$ as
\begin{align}\label{marine}
\kappa^{k,k'}(z) = q^{-\frac{n}{2}}
\frac{((zq^{n})^u; q^{k k'})_\infty}
{((-q)^r (zq^n)^u;q^{k k'})_\infty}, \qquad r=\min(k,k'),\;\;u=\max(k,k'),
\end{align}
which is the inverse of $\langle \eta_k| z^{\bf h}\ok^n | \eta_{k'}\rangle$ 
calculated from (\ref{lin}).  
It leads to the normalization
\begin{align*}
K^{k,k'}(z) |{\bf e}_1+\cdots + {\bf e}_l \rangle &=
(-1)^l|{\bf e}_{l+1}+\cdots + {\bf e}_n \rangle + \cdots
\qquad (0 \le l \le n,\; 1 \le k,k' \le 2).
\end{align*}

\section{Concluding remarks}\label{sec:end}

In this paper we have proposed the quantized reflection equation (\ref{grk})
and presented a solution in terms of the $q$-boson values $L$ and $K$ matrices 
in (\ref{ngh1}) and (\ref{ngh2}) and most notably 
the intertwiner of $A_q(sp_4)$ module known as the 3D $\Km$ \cite{KO1}
in (\ref{Kact})--(\ref{rest}).
From its $n$-concatenation the pair
$(S^{\mathrm{tr}}(z), K^{\mathrm{tr}}(z))$ 
is constructed by the trace reduction in
(\ref{obata1}), (\ref{misato}) and  
$(S^{s,s'}(z), K^{k,k'}(z))$ by the boundary vector reduction
in (\ref{obata2}), (\ref{sizka}).
They are all expressed in the matrix product form and yield new solutions to 
the reflection equation as in (\ref{misakiB}) and (\ref{aoi}).
Our boundary vector reduction 
is based on the yet conjectural property (\ref{syki})\footnote{
Besides (\ref{syki}), the relevant reflection equation (\ref{aoi}) have been verified 
for $n=2$ and for many examples from $n=3$.}.

The matrices $S(z) = S^{\mathrm{tr}}(z)$ and $S^{s,s'}(z)$ satisfy the 
Yang-Baxter equation by themselves.
In fact, as detailed in Appendix \ref{app:str}, 
they are quantum $R$ matrices for 
finite dimensional representations of 
quantum affine algebras $U_p(\mathfrak{g})$ with 
$\mathfrak{g} = A^{(1)}_{n-1}, B^{(1)}_n, \tilde{B}^{(1)}_n, D^{(1)}_n$ and 
$D^{(2)}_{n+1}$.
To summarize our solutions, we list 
$\mathfrak{g}$, the associated $S(z)$ in (\ref{sar1})--(\ref{sar5})
and those $K(z)$'s that can be paired 
with the $S(z)$  to jointly constitute a solution $(S(z),K(z))$  to the 
reflection equation. 
\begin{table}[h]
\begin{tabular}{c|c|c}
$\mathfrak{g}$ & $R$ matrix & $K$ matrix \\
\hline
$A^{(1)}_{n-1}$ & $S^{\mathrm{tr}}(z)$  & $K^{\mathrm{tr}}(z)$,\\
$D^{(2)}_{n+1}$  & $S^{1,1}(z)$ & $K^{1,1}(z)$, $K^{1,2}(z)$, 
$K^{2,1}(z)$, $K^{2,2}(z)$\\
$B^{(1)}_n$ & $S^{2,1}(z)$ & $K^{2,1}(z)$, $K^{2,2}(z)$\\
$\tilde{B}^{(1)}_n$ & $S^{1,2}(z)$ & $K^{1,2}(z)$,  $K^{2,2}(z)$\\
$D^{(1)}_n$ & $S^{2,2}(z)$ & $ K^{2,2}(z)$
\end{tabular}
\end{table}

The Yang-Baxter equation has been established 
for $S^{\mathrm{tr}}(z)$ in \cite{BS}
and $S^{s,s'}(z)$ in \cite{KS}.
In this paper we have constructed 
$ K^{\mathrm{tr}}(z), K^{k,k'}(z)$ and proved the reflection equation for 
$(S^{\mathrm{tr}}(z), K^{\mathrm{tr}}(z))$.
We have also shown that 
the reflection equation for $(S^{s,s'}(z), K^{k,k'}(z))$ in the above table 
holds provided that the conjecture (\ref{syki}) is valid.
 
This paper achieves the first systematic solutions to the reflection equation
by a method of matrix product connected to the 3D integrability. 
It suggests a number of future problems.

(i) The $q$-boson algebra (\ref{ngh1}) admits families of 
automorphisms as 
\begin{alignat*}{3}
\iota_+: \;\ap &\mapsto  u \ap \ok^{\nu},
&\quad 
\am &\mapsto  u^{-1} \ok^{-\nu}\am,
&\quad 
\ok &\mapsto  \pm  \ok,\\
\iota_- :\; \ap &\mapsto  -uq \am\ok^{\nu-1},
&\quad 
\am &\mapsto  u^{-1} \ok^{-\nu-1}\ap,
&\quad \,
\ok &\mapsto \pm  \ok^{-1}
\end{alignat*}
containing parameters $u \in \C^\times$ and $\nu \in \Z$.
The same holds also for (\ref{ngh2}) by replacing $q$ by $q^3$.
It deserves an investigation how this degrees of freedom can possibly lead to 
a generalization of the results in this paper.

(ii) Prove (\ref{syki}) and 
more generally
classify the eigenvectors of $\Rm$ and $\Km$ which are factorized as in 
(\ref{serge}) and (\ref{syki}).
Such vectors will serve as boundary vectors 
to produce further solutions to the Yang-Baxter and the reflection equations.  

(iii) Study the commuting transfer matrices with boundary associated with 
the solutions in this paper.
The routine construction of the double row transfer matrices 
acquires 3D interpretation. 
They are actually double {\em layer}
transfer matrices with boundary where the rank $n$ 
specifies a length of the layer in one direction.

(iv) Explore further solutions or versions of the quantized reflection equation.
For instance it is natural to consider a counterpart of (\ref{hrk}) in which 
$V= \C^2$ is replaced by $F_{q^2}$.
It will generate a large family of matrix product 
solutions to the reflection equation. 
The resulting systems are expected to possess rich contents
both in physics and mathematics related to special functions, combinatorics 
in the crystal limit $q\rightarrow 0$,  
stochastic processes (cf. \cite{DEHP,KMO2}) and so forth.

\appendix 
\section{Explicit form of 3D $\Rm$ and 3D $\Km$}\label{app:rk}

It suffices to impose (\ref{sakra1}) for the generators 
$g=t_{ij}$ with $1 \le i,j \le 3$.
The resulting nine equations read as
\begin{equation}\label{recR}
\begin{split}
&\hat{\Rm}(\apm \otimes \ok \otimes \ichi) = 
(\apm \otimes  \ichi \otimes \ok + \ok \otimes \apm \otimes \amp )\hat{\Rm},\\
&\hat{\Rm}(\ichi \otimes \ok \otimes \apm) = 
(\ok \otimes  \ichi \otimes \apm + \amp \otimes \apm \otimes \ok)\hat{\Rm},\\
&\hat{\Rm}(\ichi \otimes \apm \otimes \ichi) = 
(\apm \otimes \ichi \otimes \apm - \ok \otimes \apm \otimes \ok)\hat{\Rm},\\
&\hat{\Rm}(\ap\otimes \am\otimes \ap-\ok \otimes \ichi \otimes \ok)
=(\am\otimes \ap\otimes \am-\ok \otimes \ichi \otimes \ok)\hat{\Rm},\\
&[\hat{\Rm},\ok \otimes \ok \otimes \ichi] 
= [\hat{\Rm}, \ichi\otimes \ok \otimes \ok] = 0.
\end{split}
\end{equation}
These are analogue of (\ref{yuna1})--(\ref{yuna2}) for $\Km$ and 
yield recursion relations on the matrix elements of $\Rm$.
With the normalization (\ref{sakra2}) the solution is unique and given by
\begin{align}
&\hat{\Rm}(|i\rangle \otimes |j\rangle \otimes |k\rangle) = 
\sum_{a,b,c\ge 0} \hat{\Rm}^{a,b,c}_{i,j,k}
|a\rangle \otimes |b\rangle \otimes |c\rangle,\label{Rabc}\\
&\hat{\Rm}^{a,b,c}_{i,j,k} =\delta^{a+b}_{i+j}\delta^{b+c}_{j+k}
\sum_{\lambda+\mu=b}(-1)^\lambda
q^{i(c-j)+(k+1)\lambda+\mu(\mu-k)}
\frac{(q^2)_{c+\mu}}{(q^2)_c}
\binom{i}{\mu}_{\!\!q^2}
\binom{j}{\lambda}_{\!\!q^2},\label{Rex}
\end{align}
where $\delta^j_{k}=\theta(j=k)$ just to save the space.
The sum (\ref{Rex}) is over $\lambda, \mu \in \Z_{\ge 0}$ 
satisfying $\lambda+\mu=b$, which is also bounded by the 
condition $\mu\le i$ and $\lambda \le j$.
For instance, the following is the list of all the nonzero 
$\hat{\Rm}^{a,b,c}_{3,1,2}$:
\begin{alignat*}{2}
\hat{\Rm}^{1,3,0}_{3,1,2} &=  -q^2 (1 - q^4) (1 - q^6),\quad &
\hat{\Rm}^{2,2,1}_{3,1,2} &=  (1 + q^2) (1 - q^6) (1 - q^2 - q^6),\\
\hat{\Rm}^{4,0,3}_{3,1,2} &=  q^6, &
\hat{\Rm}^{3,1,2}_{3,1,2} &=  -q^2 (-1 - q^2 + q^6 + q^8 + q^{10}).
\end{alignat*}
From (\ref{Rex}) 
we see $\hat{\Rm}^{a,b,c}_{i,j,k}\in q^{\xi}\Z[q^2]$, where $\xi=0,1$ is 
specified by $\xi \equiv (a-j)(c-j)$ mod 2.
See \cite[Sec.2]{KO1} for further properties.

Let us turn to an explicit formula for the 3D $\Km$ which belongs to
$\mathrm{End}(F_{q^2} \otimes F_{q}
\otimes F_{q^2}\otimes F_{q})$. 
See (\ref{kyuki}).
We set
\begin{align}\label{Kact}
\Km(|i\rangle \otimes |j\rangle \otimes |k\rangle \otimes |l\rangle) = 
\sum_{a,b,c,d \ge 0} \Km^{a, b, c, d}_{i, j, k, l}
|a\rangle \otimes |b\rangle \otimes |c\rangle \otimes |d\rangle.
\end{align}
From (\ref{yuna1})--(\ref{yuna2}) and the normalization (\ref{tsgmi}) 
the matrix element is uniquely determined \cite[Th.2.4]{KO1} as 
\begin{align}
&\mathscr{K}^{a, b, c, d}_{i, j, k, l} =
\delta^{a+b+c}_{i+j+k}\,
\delta^{b+2c+d}_{j+2k+l}\,\frac{(q^4)_i}{(q^4)_a}
\sum_{\alpha,\beta,\gamma}
\frac{(-1)^{\alpha+\gamma}}{(q^4)_{c-\beta}}q^{\phi_1}
\nonumber\\
&\qquad \times
\mathscr{K}^{i, j+k-\alpha-\beta-\gamma ,
0, l+k-\alpha-\beta-\gamma}_{a, b+c-\alpha-\beta-\gamma,0, 
c+d-\alpha-\beta-\gamma}
\left\{{k, c-\beta,  j+k-\alpha-\beta, k+l-\alpha-\beta 
\atop
\alpha, \beta, \gamma, b-\alpha, d-\alpha, 
k-\alpha-\beta, c-\beta-\gamma}\right\},
\label{bd}\\
&\phi_1 = \alpha(\alpha+2c-2\beta-1)
+(2\beta-c)(b+c+d)+\gamma(\gamma-1)-k(j+k+l), \nonumber
\end{align}
where the sum is over $\alpha, \beta, \gamma \in \Z_{\ge 0}$.
The special case 
$\mathscr{K}^{\bullet,\bullet,0,\bullet}_{\bullet,\bullet,0,\bullet}$ 
appearing in the sum is given by
\begin{align}
&\mathscr{K}^{a, b, 0, d}_{i, j, 0, l}=
\delta^{a+b}_{i+j}\,
\delta^{b+d}_{j+l}\sum_{\lambda}(-1)^{b+\lambda}
\frac{(q^4)_{a+\lambda}}{(q^4)_a}q^{\phi_2}
\left\{{j,l \atop \lambda, l-\lambda, b-\lambda, j-b+\lambda}\right\},
\label{rest}\\
&\phi_2 = (i+a+1)(b+l-2\lambda)+b-l,\nonumber
\end{align}
where the sum is over $\lambda \in \Z_{\ge 0}$.
In (\ref{bd}) and (\ref{rest}) we have used the notation
\begin{align*}
\left\{i_1,\ldots, i_r \atop j_1, \ldots, j_s\right\} =
\begin{cases} 
\frac{\prod_{k=1}^r(q^2)_{i_k}}{\prod_{k=1}^s(q^2)_{j_k}} & 
\forall i_k, j_k \in \Z_{\ge 0},\\
0 & \text{otherwise}
\end{cases}
\end{align*}
without requiring $\sum_{k=1}^r i_k = \sum_{k=1}^rj_k$.
Due to the definition of the symbol
$\left\{ \cdots \right\}$  
the sums $\sum_{\alpha,\beta,\gamma}$ in (\ref{bd})
and $\sum_\lambda$ in (\ref{rest}) are both finite ones.
It has been shown \cite[Th.3.5]{KO1} that
$K^{a,b,c,d}_{i,j,k,l} \in q^\eta \Z[q^2]$ holds, where 
$\eta=0,1$ is specified by $\eta \equiv bd+jl \mod 2$. 
For instance the following is 
the list of all the nonzero $\Km^{1,1,1,1}_{i, j, k, l}$:
\begin{align*}
\Km^{1, 1, 1, 1}_{0, 2, 1, 0} &=q^5 (1 + q^2) (1 - q^2 -q^6),\\ 
\Km^{1, 1, 1, 1}_{0, 3, 0, 1}
&=-q^2 (1 - q^6) (1 - q^2 - q^4 - q^6 - q^8),\\ 
\Km^{1, 1, 1, 1}_{1, 0, 2, 0} &=-q (1 + q^2) (1 + q^4) (
   1 - q^4 + q^{10}), \\
\Km^{1, 1, 1, 1}_{1, 1, 1, 1} &=(1 - q^4 - q^8) (
   1 - q^2 - q^4 + q^8 + q^{10}),\\
\Km^{1, 1, 1, 1}_{1, 2, 0, 2} &=-q^5 (1 + q^2) (1 - q^4)
(2 - q^2 + q^4 - 2 q^6 - q^{10}),\\
\Km^{1, 1, 1, 1}_{2, 0, 1, 2} 
&=q (1 + q^2) (1 - q^8) (1 - q^4 - q^8 + q^{10} + q^{14}),\\
\Km^{1, 1, 1, 1}_{2, 1, 0, 3} 
&=q^2 (1 + q^2) (1 + q^4) (1 - q^6)^2 (1 - q^2 - q^8), \\
\Km^{1, 1, 1, 1}_{3, 0, 0, 4}&=q^5 (1 + q^2) (1 + q^4) (1 - q^6) 
(1 - q^8) (1 - q^{12}).
\end{align*}
See \cite[Sec.3]{KO1} for further properties.

\section{$S^{\mathrm{tr}}(z)$ and 
$S^{s,s'}(z)$ as quantum $R$ matrices}\label{app:str}

The solutions to the Yang-Baxter equation
$S^{\mathrm{tr}}(z)$ 
(\ref{obata1}), (\ref{nami})  and 
$S^{s, s'}(z)$ (\ref{obata2})
are identified with the quantum $R$ matrices 
which are characterized by the commutativity with 
quantum groups.

\subsection{Quantum affine algebras}
Consider the 
Drinfeld-Jimbo quantum affine algebra (without derivation)
$U_p(A^{(1)}_n ), U_p(D^{(2)}_{n+1}),
U_p(B^{(1)}_n), U_p(\tilde{B}^{(1)}_n)$ and $U_p(D^{(1)}_n)$.
They are Hopf algebras 
generated by $e_i, f_i, k^{\pm 1}_i\, (0 \le i \le n)$ satisfying 
\begin{equation}\label{uqdef}
\begin{split}
&k_i k^{-1}_i = k^{-1}_i k_i = 1,\;\; [k_i, k_j]=0,\;\;
k_ie_jk^{-1}_i = p_i^{a_{ij}}e_j,\;\; 
k_if_jk^{-1}_i = p_i^{-a_{ij}}f_j,\;\;
[e_i, f_j]=\delta_{ij}\frac{k_i-k^{-1}_i}{p_i-p^{-1}_i}
\end{split}
\end{equation}
together with the $p$-Serre relations \cite{D,Ji}.
We employ the coproduct $\Delta$ of the form 
\begin{align}\label{Del}
\Delta k^{\pm 1}_i = k^{\pm 1}_i\otimes k^{\pm 1}_i,\quad
\Delta e_i = 1\otimes e_i + e_i \otimes k_i,\quad
\Delta f_i = f_i\otimes 1 + k^{-1}_i\otimes f_i.
\end{align}
We follow the convention in \cite{Kac} to determine the
Cartan matrix $(a_{ij})_{0 \le i,j \le n}$ from the Dynkin diagrams\footnote{
The solutions to the Yang-Baxter equation $S^{s,s'}(z)$ which will be 
linked in (\ref{sar1})--(\ref{sar5}) are also shown. 
One observes that 
the end shape of the Dynkin diagrams is reflected in $s, s'$, namely, the 
boundary vectors $\langle \chi_s |$, $|\chi_{s'}\rangle$ in (\ref{xk}).
}:

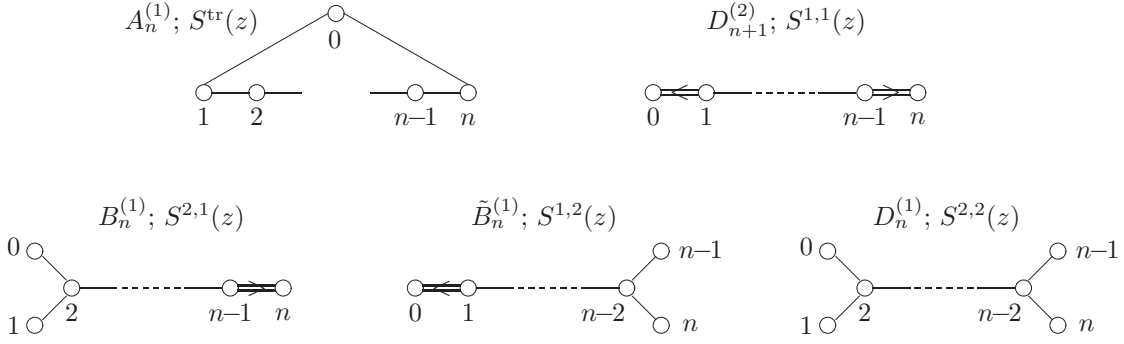
\begin{picture}(400, 150)(-31,-8)

\put(30,94){
\put(-10,24){$A^{(1)}_{n}; \,S^{\mathrm{tr}}(z)$}
\drawline(20,3)(67,30)
\put(70,30){\circle{6}}
\drawline(73,30)(120,3)
\multiput( 20,0)(20,0){2}{\circle{6}}
\multiput(100,0)(20,0){2}{\circle{6}}
\multiput(23,0)(20,0){2}{\line(1,0){14}}
\put(83,0){\line(1,0){14}}\put(103,0){\line(1,0){14}}
\put(20,-6){\makebox(0,0)[t]{$1$}}
\put(40,-6){\makebox(0,0)[t]{$2$}}
\put(100,-6){\makebox(0,0)[t]{$n\!\! -\!\! 1$}}
\put(120,-8){\makebox(0,0)[t]{$n$}}
\put(67,17){0}
}


\put(220,94){
\put(20,24){$D^{(2)}_{n+1}; \,S^{1,1}(z)$}
\multiput( 0,0)(20,0){2}{\circle{6}}
\multiput(80,0)(20,0){2}{\circle{6}}
\put(23,0){\line(1,0){14}}
\put(62.5,0){\line(1,0){14}}
\multiput(2.85,-1)(0,2){2}{\line(1,0){14.3}} 
\multiput(82.85,-1)(0,2){2}{\line(1,0){14.3}} 
\multiput(39,0)(4,0){6}{\line(1,0){2}} 
\put(10,0.2){\makebox(0,0){$<$}}
\put(90,0.2){\makebox(0,0){$>$}}
\put(0,-6){\makebox(0,0)[t]{$0$}}
\put(20,-6){\makebox(0,0)[t]{$1$}}
\put(80,-6){\makebox(0,0)[t]{$n\!\! -\!\! 1$}}
\put(100,-7.8){\makebox(0,0)[t]{$n$}}
}

\put(-20,20){
\put(30,24){$B^{(1)}_{n}; \, S^{2,1}(z)$}
\put(6,14){\circle{6}}\put(6,-14){\circle{6}}
\put(20,0){\circle{6}}
\multiput(80,0)(20,0){2}{\circle{6}}

\put(23,0){\line(1,0){14}}
\put(62.5,0){\line(1,0){14}}
\put(18,3){\line(-1,1){9}} \put(18,-3){\line(-1,-1){9}}

\multiput(82.85,-1)(0,2){2}{\line(1,0){14.3}} 
\multiput(39,0)(4,0){6}{\line(1,0){2}} 
\put(90,0){\makebox(0,0){$>$}}
\put(-2,19){\makebox(0,0)[t]{$0$}}
\put(-2,-11){\makebox(0,0)[t]{$1$}}
\put(20,-6){\makebox(0,0)[t]{$2$}}
\put(80,-6){\makebox(0,0)[t]{$n\!\! -\!\! 1$}}
\put(100,-7.8){\makebox(0,0)[t]{$n$}}
}


\put(130,20){
\put(21,24){$\tilde{B}^{(1)}_{n}; \, S^{1,2}(z)$}
\put(93,14){\circle{6}}\put(93,-14){\circle{6}}
\multiput(0,0)(20,0){2}{\circle{6}}
\put(80,0){\circle{6}}
\put(23,0){\line(1,0){14}}
\put(63,0){\line(1,0){14}}

\put(82,3){\line(1,1){9}}\put(82,-3){\line(1,-1){9}}

\multiput(2.85,-1)(0,2){2}{\line(1,0){14.3}} 
\multiput(39,0)(4,0){6}{\line(1,0){2}} 
\put(10,0){\makebox(0,0){$<$}}
\put(108,18){\makebox(0,0)[t]{$n\!\! -\!\! 1$}}
\put(0,-6){\makebox(0,0)[t]{$0$}}
\put(20,-6){\makebox(0,0)[t]{$1$}}
\put(71,-6){\makebox(0,0)[t]{$n\!\! -\!\! 2$}}
\put(104,-12){\makebox(0,0)[t]{$n$}}
}


\put(280,20){
\put(23,24){$D^{(1)}_{n}; \, S^{2,2}(z)$}
\put(6,14){\circle{6}}\put(6,-14){\circle{6}}
\put(20,0){\circle{6}}
\put(80,0){\circle{6}}
\put(93,14){\circle{6}}\put(93,-14){\circle{6}}

\put(18,3){\line(-1,1){9}} \put(18,-3){\line(-1,-1){9}}
\put(23,0){\line(1,0){14}}
\multiput(39,0)(4,0){6}{\line(1,0){2}} 
\put(62.5,0){\line(1,0){14}}
\put(82,3){\line(1,1){9}}\put(82,-3){\line(1,-1){9}}

\put(-2,19){\makebox(0,0)[t]{$0$}}
\put(-2,-11){\makebox(0,0)[t]{$1$}}
\put(20,-6){\makebox(0,0)[t]{$2$}}
\put(71,-6){\makebox(0,0)[t]{$n\!\! -\!\! 2$}}
\put(108,18){\makebox(0,0)[t]{$n\!\! -\!\! 1$}}
\put(104,-12){\makebox(0,0)[t]{$n$}}

}

\end{picture}

\noindent
Here the affine Lie algebra $\tilde{B}^{(1)}_n$ is just 
$B^{(1)}_n$ but only with different enumeration of the nodes as shown above.
We keep it for uniformity of the description although.
The constants $p_i\, (0 \le i \le n)$ in (\ref{uqdef}) are all 
taken as $p_i=p^2$ except the following:
\begin{align*}
p_0 = p_n = p \;\;\text{for }D^{(2)}_{n+1},\qquad
p_n = p \;\; \text{for }B^{(1)}_{n},\qquad
p_0 = p \;\; \text{for }\tilde{B}^{(1)}_{n}.
\end{align*}
Thus for instance in $U_p(D^{(2)}_{n+1})$, 
one has $a_{01}=-2, a_{10}=-1$ 
and $k_0e_1 = p^{-2} e_1 k_0$, $k_1 e_0 = p^{-2}e_0k_1$
and $k_1 e_1 = p^4 e_1k_1$.
Forgetting the $0$-th node in the Dynkin diagrams yields the 
classical subalgebras 
$U_p(A_n) \subset U_p(A^{(1)}_n)$,
$U_p(B_n) \subset U_p(D^{(2)}_{n+1})$,
$U_p(D_n) \subset U_p(\tilde{B}^{(1)}_{n})$ and 
$U_p(D_n) \subset U_p(D^{(1)}_{n})$.

\subsection{Representations}

We assume that $p$ is generic throughout.
We use the notations 
$|\alb\rangle$, $|\alb|$, ${\bf V} = V^{\otimes n}$
($V = \C v_0 \oplus \C v_1 \simeq \C^2$),
${\bf V}_l$, ${\bf V}^\pm$ 
explained in the beginning of Section \ref{ss:bss} and (\ref{askB}).

First consider $U_p(A^{(1)}_{n-1})$ (rather than $U_p(A^{(1)}_{n})$).
For $0 \le l \le n$, the following map 
$\pi_{l,z}: U_p(A^{(1)}_{n-1}) \rightarrow \mathrm{End}({\bf V}_l)$
defines an irreducible representation depending on 
the spectral parameter $z$\footnote{In the left hand sides of 
(\ref{mirei1}) and (\ref{mirei2}), 
$\pi_{l,z}(g)$ and $\pi_z(g)$ are denoted by $g$ for simplicity. }:
\begin{equation}\label{mirei1}
e_j|\alb \rangle 
= z^{\delta_{j,0}}|\alb-{\bf e}_j+{\bf e}_{j+1}\rangle,\quad
f_j|\alb \rangle
= z^{-\delta_{j,0}}|\alb+{\bf e}_j-{\bf e}_{j+1}\rangle,\quad
k_j |\alb \rangle
= p^{2(\alpha_{j+1}-\alpha_{j})}|\alb\rangle,
\end{equation}
where $j \in \Z_n$.  
Any vector $|\alpha'_1,\ldots, \alpha'_n\rangle$ 
appearing in the right hand sides are 
to be understood as 0 unless 
$(\alpha'_1,\ldots, \alpha'_n) \in \{0,1\}^n$.
(This convention should also apply to (\ref{mirei2}) below.)
We call $\pi_{l,z}$ the degree-$l$ {\em antisymmetric tensor} representation
(or the $l$-th {\em fundamental} representation)
following the terminology as the representation with respect to the 
classical subalgebra $U_p(A_{n-1})$.

Let us proceed to the other algebras 
$U_p(\mathfrak{g})$ with 
$\mathfrak{g} = D^{(2)}_{n+1}, B^{(1)}_n, \tilde{B}^{(1)}_n$ and $D^{(1)}_n$
under consideration.
Define the map
$\pi_{z}: U_p(\mathfrak{g}) \rightarrow \mathrm{End}({\bf V})$
by (\ref{mirei1}) for $0 < j < n$ and the following formulas for $j=0,n$ 
depending on $\mathfrak{g}$:
\begin{alignat}{3}
D^{(2)}_{n+1}, \tilde{B}^{(1)}_{n};   \quad
e_0|\alb \rangle 
&= z|\alb+{\bf e}_1\rangle,
&\quad
f_0|\alb \rangle 
&= z^{-1}|\alb-{\bf e}_1\rangle,
&\quad
k_0|\alb \rangle 
&= p^{2\alpha_1-1}|\alb\rangle,
\nonumber\\
B^{(1)}_{n}, D^{(1)}_{n};   \quad
e_0|\alb \rangle 
&= z|\alb+{\bf e}_1+{\bf e}_2\rangle,
&\quad
f_0|\alb \rangle 
&= z^{-1}|\alb-{\bf e}_1-{\bf e}_2\rangle,
&\quad
k_0|\alb \rangle 
&= p^{2(\alpha_1+\alpha_2-1)}|\alb\rangle,
\nonumber\\
D^{(2)}_{n+1}, B^{(1)}_{n};   \quad
e_n|\alb \rangle 
&= |\alb-{\bf e}_n\rangle,
&\quad
f_n|\alb \rangle 
&= |\alb+{\bf e}_n\rangle,
&\quad
k_n|\alb \rangle 
&= p^{1-2\alpha_n}|\alb\rangle,
\nonumber\\
\tilde{B}^{(1)}_{n}, D^{(1)}_{n};   \quad
e_n|\alb \rangle 
&= |\alb-{\bf e}_{n-1}-{\bf e}_n\rangle,
&\quad
f_n|\alb \rangle 
&= |\alb+{\bf e}_{n-1}+{\bf e}_n\rangle,
&\quad
k_n|\alb \rangle 
&= p^{2(1-\alpha_n-\alpha_{n-1})}|\alb\rangle.
\label{mirei2}
\end{alignat}
For $U_p(D^{(1)}_n)$, one sees that the above action of the 
generators preserves the parity of $|\alb|$.
Therefore $\pi_z$ can be restricted to 
$\pi^\pm_z : U_p(D^{(1)}_n) \rightarrow {\bf V}^\pm$ 
(\ref{askB}).
We call $\pi_z$ ($\pi^\pm_z$ for $U_p(D^{(1)}_n)$) the {\em spin} representation
by abusing the name as a representation of the 
classical subalgebra $U_p(B_n)$ or $U_p(D_n)$.

\subsection{\mathversion{bold}Quantum $R$ matrices}

Consider $U_p=U_p(\mathfrak{g})$ with 
$\mathfrak{g}$ being any one of $A^{(1)}_{n-1}, 
D^{(2)}_{n+1}, B^{(1)}_n, \tilde{B}^{(1)}_n$ and $D^{(1)}_n$.
Let $R \in \mathrm{End}({\bf V}_l \otimes {\bf V}_m)$ 
for $\mathfrak{g}=A^{(1)}_{n-1}\, (0 \le l,m \le n)$,
$R \in \mathrm{End}({\bf V}^\sigma \otimes {\bf V}^{\sigma'})$ 
for $\mathfrak{g}=D^{(1)}_{n}\, (\sigma, \sigma' = +, -)$
and $R \in \mathrm{End}({\bf V} \otimes {\bf V})$ 
for the other $\mathfrak{g}$.
Consider the linear equation on $R$
\begin{align}\label{kokona}
\Delta'_{x,y}(g) R = R\, \Delta_{x,y}(g) \qquad \forall g \in U_p,
\end{align}
where 
$\Delta_{x,y}$ signifies the tensor product representation 
$(\pi_{l,x}\otimes \pi_{m,y})\circ \Delta$ 
for $U_p(A^{(1)}_{n-1})$,
$(\pi^\sigma_{x}\otimes \pi^{\sigma'}_{y})\circ \Delta$
for  $U_p(D^{(1)}_{n})$ and 
$(\pi_{x}\otimes \pi_{y})\circ \Delta$ for the other algebras.
Similarly $\Delta'_{x,y}$ is defined 
by replacing $\Delta$ with the opposite coproduct $\Delta'$ in $\Delta_{x,y}$.
A little inspection tells that $R$ actually depends on $x$ and $y$ only via 
the ratio $z=x/y$.
The tensor product representation 
$\Delta_{x,y}$ is irreducible for generic $x/y$, hence $R$ is determined 
uniquely up to an overall scalar.
Denote them by 
$R_{l,m}(z|A^{(1)}_{n-1})$
for $U_p(A^{(1)}_{n-1})$
and $R^{\sigma,\sigma'}(z|D^{(1)}_n)$  
for $U_p(D^{(1)}_{n})$ and 
$R(z|\mathfrak{g})$ for the other cases
$\mathfrak{g}= D^{(2)}_{n+1}, B^{(1)}_n, \tilde{B}^{(1)}_n$.
They all satisfy the Yang-Baxter equation.

For $\mathfrak{g} \neq A^{(1)}_{n-1}$, 
we introduce a slight gauge transformation 
retaining the Yang-Baxter equation:
\begin{align}\label{konoha}
\tilde{R}_\pm(z|\mathfrak{g}) &= (\mathcal{I}_\pm^{-1}\otimes 1)
R(z|\mathfrak{g})(1 \otimes \mathcal{I}_\pm).
\qquad
\mathcal{I}_\pm |\alb\rangle = (\pm \I)^{|\alb|} |\alb\rangle.
\end{align}
When $\mathfrak{g} = D^{(1)}_n$ 
this should be applied to define 
$\tilde{R}^{\sigma, \sigma'}_\pm(z|D^{(1)}_n)$ for each $(\sigma, \sigma')$.

Now the identification of 
$S^{\mathrm{tr}}(z)$ (\ref{obata1}), (\ref{nami}) and 
$S^{s,s'}(z)$ (\ref{obata2}) 
with the quantum $R$ matrices is stated as follows:
\begin{alignat}{2}
S^{\mathrm{tr}}_{l,m}(z) &= R_{l,m}(z|A^{(1)}_{n-1}) 
&\qquad  &p^2 =-q^{-2},
\label{sar1}\\
S^{1,1}(z) &= 
\tilde{R}_\pm(z|D^{(2)}_{n+1})
& &p = \pm\I q^{-1},
\label{sar2}\\
S^{2,1}(z) &= 
\tilde{R}_\pm(z|B^{(1)}_{n})
& &p = \pm\I q^{-1},
\label{sar3}\\
S^{1,2}(z) &= 
\tilde{R}_\pm(z|\tilde{B}^{(1)}_{n})
& &p = \pm\I q^{-1},
\label{sar4}\\
S^{2,2}_{\sigma,\sigma'}(z) &= 
\tilde{R}^{\sigma,\sigma'}_\pm(z|D^{(1)}_{n})
& &p^2 = -q^{-2},
\label{sar5}
\end{alignat}
where we assume that the $R$ matrices in the right hand sides have been normalized 
in parallel with (\ref{askS1}).
In (\ref{sar1}) and (\ref{sar4}), it suffices to specify $p^2$ since our 
definition of 
$U_p(A^{(1)}_{n-1}), U_p(D^{(1)}_n)$\footnote{They are
referred to as
$U_{p^2}(A^{(1)}_{n-1})$ and $U_{p^2}(D^{(1)}_n)$ in 
the usual convention.} and 
their representations contain $p$ only via $p^2$.
In particular
$\tilde{R}^{\sigma,\sigma'}_+(z|D^{(1)}_{n})
= \tilde{R}^{\sigma,\sigma'}_-(z|D^{(1)}_{n})$ holds 
because the spin representation (\ref{mirei2}) always changes 
$|\alb|$ by an even number.
Up to conventional difference 
(\ref{sar1}) was claimed in \cite{BS}.
The results (\ref{sar2}), (\ref{sar3}), (\ref{sar5}) were 
proved in \cite[Th.7.1]{KS} and 
(\ref{sar4}) was suggested in \cite[Rem.7.2]{KS}. 
The essence of the proof is to show that the matrix product forms implied by the 
left hand sides fulfill the characterization (\ref{kokona}) of the $R$ matrices.

\section{Examples}\label{app:ex}

Let us write down a few examples of 
$S^{\mathrm{tr}}(z)$,
$K^{\mathrm{tr}}(z)$,
$S^{s,s'}(z)$
and 
$K^{k,k'}(z)$ 
explicitly.

\subsection{\mathversion{bold}$S^{\mathrm{tr}}_{m,1}(z)$ 
and $S^{\mathrm{tr}}_{1,m}(z)$ with general $m, n$}

The $S^{\mathrm{tr}}_{l,m}(z)$ in (\ref{nami}) 
is an elementary example of 
quantum $R$ matrices
associated with the antisymmetric tensor representations 
as noted in (\ref{sar1}).
When $\min(l,m)=1$, its nonzero matrix elements are given as
\begin{align*}
S^{\mathrm{tr}}_{m,1}(z)^{\alb,{\bf e}_j}_{\alb,{\bf e}_j}
&= \begin{cases}
(-1)^m\frac{q^2(1-q^{2m-2}z)}{1-q^{2m+2}z}& \alpha_j = 1,\\
(-1)^{m+1} & \alpha_j=0,
\end{cases}
\\
S^{\mathrm{tr}}_{m,1}(z)_{\alb,{\bf e}_j}^{\gab,{\bf e}_k}
&= \begin{cases}
(-1)^{m+1}\frac{z(1-q^4)}{1-q^{2m+2}z}
q^{2(m-\alpha_{j+1}-\alpha_{j+2}-\cdots - \alpha_k)}
& j<k,\\
(-1)^{m+1}\frac{1-q^4}{1-q^{2m+2}z}
q^{2(\alpha_{k+1}+\alpha_{k+2}+\cdots + \alpha_j)}
& j>k,
\end{cases}
\\
S^{\mathrm{tr}}_{1,m}(z)^{{\bf e}_j, \beb}_{{\bf e}_j, \beb}
&= \begin{cases}
1 & \beta_j = 1,\\
-\frac{q^2(1-q^{2m-2}z)}{1-q^{2m+2}z} & \beta_j=0,
\end{cases}
\\
S^{\mathrm{tr}}_{1,m}(z)_{{\bf e}_j, \beb}^{{\bf e}_k, \deb}
&= \begin{cases}
\frac{1-q^4}{1-q^{2m+2}z}
q^{2(\delta_{j+1}+\delta_{j+2}+\cdots + \delta_k)}
& j<k, \\
\frac{z(1-q^4)}{1-q^{2m+2}z}
q^{2(m-\delta_{k+1}-\delta_{k+2}-\cdots - \delta_j)}
& j>k,
\end{cases}
\end{align*}
where $\alb, \beb, \gab, \deb \in \{0,1\}^n$ with 
$|\alb| =|\beb| =|\gab| =|\deb| =m$.  
The case $m=1$ corresponds to the well known $n(2n-1)$-vertex model
associated with the vector representation.
In particular the case $n=2$ is the six-vertex model in which  
$S^{\mathrm{tr}}_{1,1}(z)$ acts on the base vectors as
\begin{align*}
|ij, ij \rangle & \mapsto  |ij, ij \rangle\quad (i,j\in\{0,1\}),
\\ 
 | 01, 10\rangle  &\mapsto -\frac{q^2 (-1 + z) | 01, 10\rangle }{-1 + q^4 z} 
 + \frac{(-1 + q^4) z | 10, 01\rangle }{-1 + q^4 z},
   \\ 
   | 10, 01\rangle  &\mapsto \frac{(-1 + q^4) | 01, 10\rangle }{-1 + q^4 z} 
   - \frac{q^2 (-1 + z) | 10, 01\rangle }{-1 + q^4 z}.
\end{align*}
Here we have written $(v_1 \otimes v_0) \otimes (v_1\otimes v_1)
\in V^{\otimes 2} \otimes V^{\otimes 2}$ for example as 
$|10,11\rangle$ for simplicity. 

\subsection{\mathversion{bold} $K^{\mathrm{tr}}(z)$ for $n=2, 3$}

Denote $v_0 \otimes v_1$ by $|01\rangle$ etc.
When $n=2$, $K^{\mathrm{tr}}(z)$ acts on the base vectors as 
\begin{align*}
| 00\rangle  &\mapsto | 11\rangle ,
\qquad
| 01\rangle  \mapsto -\frac{q^{-1}(-1 + q^2) z | 01\rangle }{(-1 + z)} + | 10\rangle ,\\
| 11\rangle  &\mapsto | 00\rangle ,
\qquad
| 10\rangle  \mapsto | 01\rangle  - \frac{q^{-1}(-1 + q^2) | 10\rangle }{(-1 + z)}.
\end{align*}

When $n=3$, 
$K^{\mathrm{tr}}(z)$ acts on the base vectors as 
\begin{align*}
| 000\rangle  &\mapsto | 111\rangle ,\qquad
  | 111\rangle  \mapsto  | 000\rangle ,
 \\ 
 | 001\rangle  &\mapsto  -\frac{(-1 + q^2) z | 011\rangle }{
   q (-1 + q z)} - \frac{(-1 + q^2) z | 101\rangle }{-1 + q z} 
   +  | 110\rangle ,
  \\ 
  | 010\rangle  &\mapsto  -\frac{(-1 + q^2) z | 011\rangle }{-1 + q z} 
  + | 101\rangle  - \frac{(-1 + q^2) | 110\rangle }{q (-1 + q z)},
  \\ 
  | 011\rangle  &\mapsto  -\frac{(-1 + q^2) z | 001\rangle }{
   q (-1 + q z)} - \frac{(-1 + q^2) z | 010\rangle }{-1 + q z} + 
  | 100\rangle ,
  \\ 
  | 100\rangle  &\mapsto  
 | 011\rangle  - \frac{(-1 + q^2) | 101\rangle }{
  q (-1 + q z)} - \frac{(-1 + q^2) | 110\rangle }{-1 + q z},
  \\ 
  | 101\rangle  &\mapsto  -\frac{(-1 + q^2)z | 001\rangle }{-1 + q z} + 
  | 010\rangle  - \frac{(-1 + q^2) | 100\rangle }{q (-1 + q z)},
  \\ 
  | 110\rangle  &\mapsto  | 001\rangle  - \frac{(-1 + q^2) | 010\rangle }{
  q (-1 + q z)} - \frac{(-1 + q^2) | 100\rangle }{-1 + q z}.
\end{align*}
These formulas are consistent with (\ref{misato3}).

\subsection{\mathversion{bold}$S^{s,s'}(z)$ and $K^{k,k'}(z)$ for $n=1$}
Let us present the action of $S^{s,s'}(z)$ on $V\otimes V$.
From the parity constraint (\ref{yume4}), 
$S^{2,2}(z)$ becomes diagonal whose elements are 
already fixed by the normalization condition (\ref{askS1}).
In the remaining cases we will only cover  
$(s,s')=(1,1), (1,2)$ in view of (\ref{yume15}).
We write, for example, as
$|0,1\rangle = v_0 \otimes v_1$.
\begin{align*}
S^{1,1}(z): \;\;&|0,0\rangle \mapsto |0,0\rangle,\quad
|0,1\rangle \mapsto 
\frac{q (1 - z) |0,1\rangle}{1 + q^2 z} 
+ \frac{(1 + q^2) z |1,0\rangle}{1 + q^2 z},\\
&|1,1\rangle \mapsto |1,1\rangle,\quad
|1,0\rangle \mapsto 
\frac{(1 + q^2) |0,1\rangle}{1 + q^2 z}
+\frac{q (-1 + z) |1,0\rangle}{1 + q^2 z},\\
S^{1,2}(z): \;\;&|0,0\rangle \mapsto |0,0\rangle,\quad
|0,1\rangle \mapsto 
\frac{q (1 - z^2) |0,1\rangle}{1 + q^2 z^2} 
+ \frac{(1 + q^2) z |1,0\rangle}{1 + q^2 z^2},\\
& |1,1\rangle \mapsto |1,1\rangle,\quad
|1,0\rangle \mapsto 
\frac{(1 + q^2)z |0,1\rangle}{1 + q^2 z^2}
+\frac{q (-1 + z^2) |1,0\rangle}{1 + q^2 z^2}.
\end{align*}
So $S^{1,1}(z)$ and $S^{1,2}(z)$ 
define just six vertex models
in some gauge.

For $K^{k,k'}(z)$, we will again cover 
$(k,k')=(1,1), (1,2)$ and $(2,2)$ only by virtue of (\ref{aimi2}). 

\begin{alignat*}{2}
K^{1,1}(z): |0\rangle &\mapsto 
-\frac{q^{-\hf}(1 + q) z |0\rangle}{-1 + z} + |1\rangle,&\qquad 
|1\rangle &\mapsto -|0\rangle - \frac{q^{-\hf}(1 + q) |1\rangle}{-1 + z},
\\
K^{1,2}(z):|0\rangle &\mapsto  
-\frac{q^{-\hf}(1 + q) z |0\rangle}{-1 + z^2} + |1\rangle, &\qquad
|1\rangle &\mapsto  -|0\rangle - \frac{q^{-\hf}(1 + q) z |1\rangle}{-1 + z^2},
\\
K^{2,2}(z): |0\rangle &\mapsto |1\rangle,&\qquad |1\rangle &\mapsto -|0\rangle.
\end{alignat*}

\subsection{\mathversion{bold}$S^{1,1}(z)$ and $K^{k,k'}(z)$ for $n=2$}

We set 
$|10,11\rangle = (v_1 \otimes v_0) \otimes (v_1\otimes v_1)$ etc as before.
The $S^{1,1}(z)$ acts 
on the base vectors of $V^{\otimes 2} \otimes V^{\otimes 2}$ as follows:
\begin{align*}
|ij, ij \rangle & \mapsto  |ij, ij \rangle\quad (i,j\in\{0,1\}),
\\ 
|00, 01 \rangle
& \mapsto   -\frac{q (-1 + z) |00, 01 \rangle}{1 + q^2 z} 
+ \frac{(1 + q^2) z |01, 00 \rangle}{1 + q^2 z},
\\
 |00, 10 \rangle
& \mapsto   -\frac{q (-1 + z) |00, 10 \rangle}{1 + q^2 z} 
+ \frac{(1 + q^2) z |10, 00 \rangle}{1 + q^2 z},
\\
 |00, 11 \rangle
& \mapsto  \frac{q^2 (-1 + z) (-1 + q^2 z) |00, 11\rangle}{
(1 + q^2 z) (1 + q^4 z)} - 
\frac{q^3 (1 + q^2) (-1 + z) z |01, 10\rangle}{
(1 + q^2 z) (1 + q^4 z)}\\
& \quad- 
\frac{q (1 + q^2) (-1 + z) z |10, 01\rangle}{
(1 + q^2 z) (1 + q^4 z)} 
+ \frac{(1 + q^2) (1 + q^4) z^2 |11, 00 \rangle}{
(1 + q^2 z) (1 + q^4 z)},
\\ 
|01, 00\rangle & \mapsto  
\frac{(1 + q^2) |00, 01 \rangle}{1 + q^2 z} + 
\frac{q (-1 + z) |01, 00 \rangle}{1 + q^2 z},
\\ 
|01, 10 \rangle
& \mapsto  -\frac{q (1 + q^2) (-1 + z) |00, 11\rangle}{
(1 + q^2 z) (1 + q^4 z)} - 
\frac{q^2 (-1 + z) (-1 + q^2 z) |01, 10\rangle}{
(1 + q^2 z) (1 + q^4 z)} \\
& \quad +
\frac{(1 + q^2) z (1 + q^2 - q^2 z + q^4 z) |10, 01 \rangle}{
(1 + q^2 z) (1 + q^4 z)} 
+ \frac{q (1 + q^2) (-1 + z) z |11, 00\rangle}{(1 + q^2 z) (1 + q^4 z)},
\\
 |01, 11 \rangle
& \mapsto  -\frac{q (-1 + z) |01, 11 \rangle}{1 + q^2 z} + 
\frac{(1 + q^2) z |11, 01 \rangle}{1 + q^2 z},
\\ 
|10, 00 \rangle
& \mapsto  \frac{(1 + q^2) |00, 10 \rangle}{1 + q^2 z} + 
\frac{q (-1 + z) |10, 00 \rangle}{1 + q^2 z},
\\ 
|10, 01 \rangle& \mapsto  
-\frac{q^3 (1 + q^2) (-1 + z) |00, 11\rangle}{(1 + q^2 z) (1 + q^4 z)} 
+ \frac{(1 + q^2) (1 - q^2 + q^2 z + q^4 z) |01, 10 \rangle}{
(1 + q^2 z) (1 + q^4 z)} \\
& \quad 
- \frac{q^2 (-1 + z) (-1 + q^2 z) |10, 01\rangle}{(1 + q^2 z) (1 + q^4 z)} 
+ \frac{q^3 (1 + q^2) (-1 + z) z |11, 00\rangle}{(1 + q^2 z) (1 + q^4 z)},
\\
 |10, 11 \rangle
& \mapsto  -\frac{q (-1 + z) |10,11 \rangle}{1 + q^2 z} 
+ \frac{(1 + q^2) z |11, 10 \rangle}{1 + q^2 z},
\\
 |11, 00 \rangle
& \mapsto  \frac{(1 + q^2) (1 + q^4) |00, 11\rangle}{
(1 + q^2 z) (1 + q^4 z)} 
+ \frac{q^3 (1 + q^2) (-1 + z) |01, 10\rangle}{
(1 + q^2 z) (1 + q^4 z)} \\
&\quad + \frac{q (1 + q^2) (-1 + z) |10, 01\rangle}{
(1 + q^2 z) (1 + q^4 z)} + 
\frac{q^2 (-1 + z) (-1 + q^2 z) |11, 00\rangle}{(1 + q^2 z) (1 + q^4 z)},
\\
 |11, 01 \rangle
& \mapsto  \frac{(1 + q^2) |01, 11 \rangle}{1 + q^2 z} + 
\frac{q (-1 + z) |11, 01 \rangle}{1 + q^2 z},
\\
 |11, 10 \rangle& \mapsto  
\frac{(1 + q^2) |10, 11 \rangle}{1 + q^2 z} + 
\frac{q (-1 + z) |11, 10 \rangle}{1 + q^2 z}.
\end{align*}

$K^{1,1}(z)$ acts on the base vectors of $V^{\otimes 2}$ as follows:
\begin{align*}
|0 0\rangle & \mapsto \frac{q^{-1}(1 + q) (1 + q^2) z^2 |0 0\rangle }{(-1 + z) (-1 + q z)} 
- \frac{q^{-\hf}(1 + q) z |0 1\rangle }{(-1 + q z)} - \frac{q^{\hf}(1 + q) z |1 0\rangle }{-1 + q z} + 
  |1 1\rangle , \\ 
  |0 1\rangle & \mapsto \frac{q^{-\hf}(1 + q) z |0 0\rangle }{-1 + q z} 
  + \frac{q^{-1}(1 + q) z (1 + q - q z + q^2 z) |0 1\rangle }{(-1 + z) (-1 + q z)} 
  - |1 0\rangle  - \frac{q^{-\hf}(1 + q) |1 1\rangle }{-1 + q z}, 
  \\ 
  |1 0\rangle & \mapsto \frac{q^{\hf}(1 + q) z |0 0\rangle }{-1 + q z} 
  -|0 1\rangle  
  + \frac{q^{-1}(1 + q) (1 - q + q z + q^2 z) |1 0\rangle }{(-1 + z) (-1 + q z)} 
  - \frac{q^{\hf} (1 + q) |1 1\rangle }{-1 + q z}, 
  \\ 
  |1 1\rangle & \mapsto 
 |0 0\rangle  + \frac{q^{-\hf}(1 + q) |0 1\rangle }{-1 + q z} 
 + \frac{q^{\hf} (1 + q) |1 0\rangle }{-1 + q z} 
 + \frac{q^{-1}(1 + q) (1 + q^2) |1 1\rangle }{(-1 + z) (-1 + q z)}.
\end{align*}
$K^{1,2}(z)$ acts on the base vectors of $V^{\otimes 2}$ as follows:
\begin{align*}
|0 0\rangle & \mapsto \frac{q^{-1}(1 + q) z^2 (1 + q^2 - q^2 z^2 + q^3 z^2) |0 0\rangle }{
(-1 + z^2)(-1 + q^2 z^2)} 
- \frac{q^{-\hf}(1 + q) z |0 1\rangle }{-1 + q^2 z^2} 
- \frac{q^{\hf} (1 + q) z |1 0\rangle }{-1 + q^2 z^2} 
+ |1 1\rangle , \\ 
  |0 1\rangle & \mapsto \frac{q^{-\hf}(1 + q) z |0 0\rangle }{-1 + q^2 z^2} 
  + \frac{q^{-1}(1 + q) z^2 (1 + q^2 - q^2 z^2 + q^3 z^2) |0 1\rangle }{(-1 + z^2)(-1 + q^2 z^2)}
- |1 0\rangle  
- \frac{q^{\hf} (1 + q) z |1 1\rangle }{-1 + q^2 z^2}, 
  \\ 
  |1 0\rangle & \mapsto \frac{q^{\hf} (1 + q) z |0 0\rangle }{-1 + q^2 z^2} 
  - |0 1\rangle  
  + \frac{q^{-1}(1 + q) (1 - q + q z^2 + q^3 z^2) |1 0\rangle }{
 (-1 + z^2)(-1 + q^2 z^2)} 
 - \frac{q^{\scriptstyle{\frac{3}{2}}}(1 + q) z |1 1\rangle }{-1 + q^2 z^2}, 
  \\ 
  |1 1\rangle & \mapsto |0 0\rangle  + 
  \frac{q^{\hf}(1 + q) z |0 1\rangle }{-1 + q^2 z^2} 
 + \frac{q^{\scriptstyle{\frac{3}{2}}} (1 + q) z |1 0\rangle }{-1 + q^2 z^2} 
  + \frac{q^{-1}(1 + q) (1 - q + q z^2 + q^3 z^2) |1 1\rangle }{(-1 + z^2)(-1 + q^2 z^2)}.
\end{align*}
$K^{2,2}(z)$ acts on the base vectors of $V^{\otimes 2}$ as follows:
\begin{align*}
|0 0\rangle & \mapsto \frac{q^{-1}(-1 + q^2)z^2 |0 0\rangle }{-1 + z^2} 
+ |1 1\rangle , \qquad 
  |0 1\rangle  \mapsto \frac{q^{-1}(-1 + q^2)z^2 |0 1\rangle }{-1 + z^2} - |1 0\rangle , 
  \\ 
  |1 0\rangle & \mapsto -|0 1\rangle  + \frac{q^{-1}(-1 + q^2) |1 0\rangle }{-1 + z^2}, 
  \qquad
  |1 1\rangle  \mapsto 
 |0 0\rangle  + \frac{q^{-1}(-1 + q^2)|1 1\rangle }{-1 + z^2}.
\end{align*}

\section*{Acknowledgments}
The authors thank the organizers of 
MATRIX Program {\em Non-Equilibrium Systems and Special Functions}
at University of Melbourne (Creswick, 8 January 2018 -- 2 February 2018),
which triggered this work.
AK is supported by 
Grants-in-Aid for Scientific Research 
No.~15K13429 from JSPS.

\end{document}